\documentclass{LMCS}
\usepackage{amssymb}
\usepackage{amsmath}
\usepackage{amsthm}
\usepackage{verbatim}
\usepackage[usenames]{color}
\usepackage{hyperref}

\newtheorem{la}[thm]{Lemma}
\newtheorem{coro}[thm]{Corollary}

\theoremstyle{definition}
\newtheorem{df}[thm]{Definition}
\newtheorem{ex}[thm]{Example}   
\newtheorem{rmk}[thm]{Remark}
\newtheorem{conn}[thm]{Convention}

\newenvironment{eatab}
 {\bigskip\noindent\begin{minipage}{\textwidth}\upshape\ttfamily
  \begin{tabbing}mmm\=mmm\=mmm\=mmm\=mmm\=\kill}
 {\end{tabbing}\end{minipage}\bigskip}

\newenvironment{eqs}{\begin{eqnarray*}}{\end{eqnarray*}}

\newenvironment{lsnum}{\begin{enumerate}}{\end{enumerate}}
\newenvironment{ls}{\begin{itemize}}{\end{itemize}}

\newcommand{\notarrow}{\kern .42em\not\kern -.42em\longrightarrow}

\newcommand{\DD}{\Delta^+}
\newcommand{\E}{\mathrm{E}}
\newcommand{\Issued}{\text{Issued}}
\newcommand{\U}{\Upsilon}
\newcommand{\ans}{\dot}

\newcommand{\bld}[1]{\ensuremath{\mathbf {#1}}}

\newcommand{\dom}[1]{\ensuremath{{\text{Dom}}(#1)}}
\newcommand{\emp}{\varnothing}

\newcommand{\eps}{\varepsilon}

\newcommand{\initeq}{\unlhd}
\newcommand{\init}{\lhd}

\newcommand{\kand}{\curlywedge}
\newcommand{\kor}{\curlyvee}
\newcommand{\mst}[1]{\ensuremath{\{\kern-0.33em\{#1\}\kern-0.33em\}}}
\newcommand{\nlet}{\ensuremath{\ttt{n}\text-\ttt{let}\ }}
\newcommand{\qval}[3]{\ensuremath{\text{q-Val}(#1,#2,#3)}}
\newcommand{\ran}[1]{\ensuremath{{\text{Range}}(#1)}}
\newcommand{\restr}{\mathop{\upharpoonright}}

\newcommand{\scr}[1]{\ensuremath{\mathcal {#1}}}

\newcommand{\sq}[1]{\ensuremath{\langle#1\rangle}}

\newcommand{\ttt}[1]{\ensuremath{\mathtt {#1}}}
\newcommand{\val}[3]{\ensuremath{\text{Val}(#1,#2,#3)}}
\newcommand{\vlet}{\ensuremath{\ttt{v}\text-\ttt{let}\ }}

\renewcommand{\phi}{\varphi}
\renewcommand{\th}{\ensuremath{{}^{\text{th}}}}

\def\doi{3 (4:4) 2007}
\lmcsheading%
{\doi}
{1--35}
{}
{}
{Jun.~11, 2007}
{Nov.~\phantom{0}5, 2007}
{}   

\begin{document}

\title[Interactive Small-Step Algorithms II]{Interactive Small-Step
  Algorithms II: Abstract State Machines and the Characterization
  Theorem} 
\author[A.~Blass]{Andreas Blass\rsuper a} 
\address{{\lsuper a}Mathematics Dept.,
  University of Michigan, Ann Arbor, MI 48109, U.S.A.}
\email{ablass@umich.edu} 
\thanks{{\lsuper a}Blass was partially supported by NSF
  grant DMS--0070723 and by a grant from Microsoft Research.}

\author[Y.~Gurevich]{Yuri Gurevich\rsuper b}
\address{{\lsuper b}Microsoft Research, One Microsoft Way, Redmond, WA 98052,
U.S.A.}
\email{gurevich@microsoft.com}

\author[D.~Rosenzweig]{Dean Rosenzweig\rsuper c} 
\address{{\lsuper c}University of Zagreb, FSB, I. Lu\v ci\'ca 5, 10000 Zagreb,
  Croatia}
\email{dean@math.hr}
\thanks{{\lsuper c}Rosenzweig was partially supported by the grant 0120048 from
  the Croatian Ministry of Science and Technology and by Microsoft
  Research.} 

\author[B.~Rossman]{Benjamin Rossman\rsuper d}
\address{{\lsuper d}Computer Science Dept., M.I.T., Cambridge, MA 02139, U.S.A.}
\email{brossman@mit.edu}


\begin{abstract}
In earlier work, the Abstract State Machine Thesis --- that
arbitrary algorithms are behaviorally equivalent to abstract state
machines --- was established for several classes of algorithms,
including ordinary, interactive, small-step algorithms.  This was
accomplished on the basis of axiomatizations of these classes of
algorithms. In a companion paper \cite{ga1} the axiomatisation was
extended to cover interactive small-step algorithms that are not
necessarily ordinary. This means that the algorithms (1)~can
complete a step without necessarily waiting for replies to all
queries from that step and (2)~can use not only the environment's
replies but also the order in which the replies were received.  In
order to prove the thesis for algorithms of this generality, we
extend here the definition of abstract state machines to incorporate
explicit attention to the relative timing of replies and to the
possible absence of replies. We prove  the characterization theorem
for extended ASMs with respect to general algorithms as axiomatised
in \cite{ga1}.
\end{abstract}

\dedicatory{While this paper was being revised, we received the sad
  news of the death of our co-author,\\ Dean Rosenzweig.  We dedicate
  this paper to his memory.\\
\hfill Andreas Blass, Yuri Gurevich, Benjamin Rossman}

\maketitle
\vfill

\section{Introduction}
\label{intro}

Traditional models of computation, like the venerable Turing machine,
are, despite the Church-Turing thesis, rather distant intuitively from
many of the concerns of modern computing.  Graphical user interfaces,
parallel and distributed computing, communication and security
protocols, and various other sorts of computation do not easily fit
the traditional picture of computing from input strings to output
strings.  Abstract state machines (ASMs) were introduced for the
purpose of modeling algorithms at their natural level of abstraction,
as opposed to the far lower level of abstraction usually needed by a
Turing machine model.  That ASMs fulfill their purpose was at first an
empirical fact, supported by numerous case studies, not only of
algorithms in the usual sense but also of whole programming languages
and of hardware; see \cite{web} for many examples.  The Abstract State
Machine Thesis, first proposed in \cite{G64.5} and then
elaborated in \cite{G92,G103}, asserts that every algorithm is
equivalent, on its natural level of abstraction, to an abstract state
machine.  Beginning in \cite{seqth} and continuing in \cite{parth},
\cite{oa1}, \cite{oa2}, and \cite{oa3}, the thesis has been proved for
various classes of algorithms.  In each case, the class of algorithms
under consideration was defined by postulates describing, in very
general terms, the nature of the algorithms, and in each case the main
theorem was that all algorithms of this class are equivalent, in a
strong sense, to ASMs.

The thesis was proved first, in \cite{seqth}, for the class of
algorithms that are sequential (i.e., proceed in discrete steps and do
only a bounded amount of work per step) and do not interact with the
external environment within steps.  (The environment is allowed to
intervene between steps to change the algorithm's state.)  

Subsequent work extended the result in two directions.  Parallel
algorithms, in which a bound on work per step applies to each
processor but not to the algorithm as a whole, were treated in
\cite{parth} but still without intrastep interaction with the
environment.  In \cite{oa1, oa2, oa3}, intrastep interaction was added
to sequential computation, subject to a restriction to ``ordinary''
interaction, and the ASM thesis was proved for the resulting class of
algorithms.  In both of these directions, the standard syntax of ASMs,
as presented in \cite{G103}, was adequate, with only very minor
modifications.

In the present paper and its companion paper \cite{ga1}, we continue
this tradition, now removing the restriction to ordinary interaction.
That is, we propose the postulates in \cite{ga1} as a general
description of sequential algorithms interacting with their
environments, and we show in the present paper that all algorithms
that satisfy the postulates are behaviorally equivalent, in a strong
sense, to ASMs.

There is, however, an important difference between this work and the
earlier proofs of the ASM thesis.  The traditional ASM syntax and
semantics from \cite{G103} are no longer adequate.  They require a
significant extension, allowing an ASM program (1)~to refer to the
order in which the values of external functions are received from the
environment and (2)~to declare a step complete even if not all
external function values have been determined.  Neither of these two
possibilities was permitted by the postulates defining ``ordinary
algorithm'' in \cite{oa1}.  

In \cite{ga1}, we presented postulates that permit both of these
possibilities, and we argued that these postulates capture the general
concept of sequential, interactive algorithm.  In the present paper,
we extend the syntax and semantics of abstract state machines so that
non-ordinary algorithms become expressible.  The main contributions of
this paper are
\begin{ls}
\item syntax and semantics for ASMs incorporating interaction that need
  not be ordinary,
\item verification that ASMs satisfy the postulates of \cite{ga1}, and
\item proof that every algorithm satisfying the postulates is equivalent,
  in a strong sense, to an ASM.
\end{ls}

Most design decisions about the syntax and semantics of general
interactive ASMs were guided, and often forced, by the axiomatisation
of appropriate algorithms in the companion paper \cite{ga1}.
Sections~\ref{sec:asm} and \ref{asm-sem}, defining the syntax and semantics of
interactive ASMs are self-contained and could in principle be read
independently of \cite{ga1}, but we refer the reader to \cite{ga1},
and sometimes also to \cite{oa1,oa2,oa3}, for extensive discussion,
motivation and justification of some of the choices made, as well as
the relation to other work.  Sections \ref{sec:asms-are-algos} and
\ref{sec:thesis}, relating the ASMs of section \ref{sec:asm} to
algorithms as axiomatized in \cite{ga1}, use the definitions and
results of \cite{ga1}.  We presume that the reader has a copy of the
companion paper \cite{ga1} available, but, as an aid to intuition, we
summarize briefly the main content of the postulates.

The \emph{states} of an algorithm are structures for a finite
vocabulary $\Upsilon$, and certain states are designated as
\emph{initial states}.  The algorithm's interaction with the
environment (during a step) is given by a \emph{history}, which
consists of a function sending the algorithm's queries to the
environment's answers, together with a linear pre-order telling in
what order the answers were received.  The algorithm tells what
queries are to be issued, on the basis of the state and the past
history.  On the same basis, it also tells whether the current step is
ended; if so, it tells whether the step has succeedd or failed, and in
the case of success it tells how the state is to be updated.  The
updating changes the interpretations of some of the function symbols,
but it does not affect the base set.  All of the preceding aspects of
the algorithm are required to be invariant under isomorphism of
$\Upsilon$-structures.  Finally, the ``small-step'' property of the
algorithm is ensured by a postulate saying that the queries to be
issued, the decisions about ending the step and about success, and the
updates depend only on the history plus a specific finite part of the
state.  For the technical details of the formulation of the
postulates, we refer to \cite[Section~3]{ga1}.

\section{Interactive Small-Step ASMs: Syntax}   \label{sec:asm}

Ordinary interactive small-step ASMs are defined in \cite{oa2}.  In the
companion paper \cite{ga1}, we axiomatized general interactive
small-step algorithms.  In this and the next sections, we define general
interactive small-step ASMs.  This new ASM model is an extension of the
ASM model in \cite{oa2}.  The extension incorporates capabilities for
taking into account the order of the environment's replies and for ending
a step before all queries have been answered.  We repeat here, for the
sake of completeness, some definitions from \cite{oa2,ga1}, but we do not
repeat the detailed discussion and motivation for these definitions.  We
provide detailed discussion and motivation for those aspects of the
present material that go beyond what was in \cite{oa2,ga1}.

In this section we describe the syntax of ASM programs, accompanied
with some intuitive indications of their semantics.  Precise semantics
is given in the next section.

\subsection{Vocabularies}

An ASM has a finite vocabulary $\U$ complying with the following
convention, exactly as required for interactive small-step algorithms
in \cite{ga1}.

\begin{conn}   \label{vocab}
\mbox{}
  \begin{ls}
    \item A vocabulary $\U$ consists of function symbols with
    specified arities.
    \item Some of the symbols in $\U$ may be marked as
    \emph{static}, and some may be marked as \emph{relational}.
    Symbols not marked as static are called \emph{dynamic}.
    \item Among the symbols in $\U$ are the logic names: nullary
    symbols \ttt{true}, \ttt{false}, and \ttt{undef}; unary
    \ttt{Boole}; binary equality; and the usual propositional
    connectives.  All of these are static and all but \ttt{undef} are
    relational.
    \item An $\Upsilon$-structure consists of a nonempty base set and
      interpretations of all the function symbols as functions  on
      that base set.
    \item In any $\U$-structure, the interpretations of \ttt{true},
    \ttt{false}, and \ttt{undef} are distinct.
    \item In any $\U$-structure, the interpretations of relational
    symbols are functions whose values lie in
    $\{\ttt{true}_X,\ttt{false}_X\}$.
  \item In any $\U$-structure $X$, the interpretation of
    \ttt{Boole} maps $\ttt{true}_X$ and $\ttt{false}_X$ to
    $\ttt{true}_X$ and everything else to $\ttt{false}_X$.
    \item In any $\U$-structure $X$, the interpretation of
    equality maps pairs of equal elements to $\ttt{true}_X$ and all
    other pairs to $\ttt{false}_X$.
    \item In any $\U$-structure $X$, the propositional connectives
    are interpreted in the usual way when their arguments are in
    $\{\ttt{true}_X,\ttt{false}_X\}$, and they take the value
    $\ttt{false}_X$ whenever any argument is not in
    $\{\ttt{true}_X,\ttt{false}_X\}$.
    \item We may use the same notation $X$ for a structure and its
    base set.
\item We may omit subscripts $X$, for example from \ttt{true} and
  \ttt{false}, when there is no danger of confusion.\qed
  \end{ls}
\end{conn}

In addition, the ASM
has an
\emph{external vocabulary} $\E$, consisting of finitely many
\emph{external function symbols}\footnote{The symbol $\E$ for the
external vocabulary is the Greek capital epsilon, in analogy with the
Greek capital upsilon $\U$ for the algorithm's vocabulary.}.
These symbols are used syntactically exactly like the symbols from
$\U$, but their semantics is quite different.  If $f$ is an
$n$-ary external function symbol and $\bld a$ is an $n$-tuple of
arguments from a state $X$, then the value of $f$ at $\bld a$ is not
stored in the state but is obtained from the environment as the reply
to a query.

\begin{rmk}
The ASM syntax of \cite{oa2} included commands of the form
$\ttt{Output}_l(t)$
where $t$ is a term and $l$ is a so-called output
label.  These commands produced an outgoing message, regarded as a
query with an automatic reply ``OK.''  In the present paper, we shall
include commands for issuing the queries associated with external
function calls even when the reply might not be used in the evaluation
of a  term.  These \ttt{issue} commands subsume the older \ttt{Output}
commands, so we do not include the latter in our present syntax.  This
is why the preceding paragraph introduces only the external vocabulary
and not an additional set of output labels.  Note in this connection
that the simulation of ordinary interactive small-step algorithms by
ASMs in \cite{oa3} did not use \ttt{Output} rules.
\qed\end{rmk}

\begin{conn}
  Note that by Convention~\ref{vocab} only function symbols in
  $\U$ admit two sorts of markings.   They can be either static
  or dynamic and they can be relational or not.  \emph{No such
  markings} are applied to the external function symbols.  All symbols
  in $\E$ are considered static and not relational.
\qed\end{conn}

\begin{rmk}
  In this convention, ``static'' does not mean that the values of
  external functions cannot change; it means that the algorithm cannot
  change them, although the environment can.  External functions
  cannot be the subject of updates in an ASM program, and in this
  respect they have the same syntax as static function symbols from
  $\U$.

We do not declare any external function symbols to be relational
because such a declaration would, depending on its semantical
interpretation, lead to one of two difficulties.

One possibility would be to demand that the queries resulting from
relational external functions get replies that are appropriate values
for such functions, namely only \ttt{true} and \ttt{false}.  This
imposes a burden on the environment, and a fairly complicated one,
since it may not be evident, by inspection of a query, what external
function symbol produced it (see the discussion of templates below).
We prefer in this paper to keep the environment unconstrained.

A second possibility for handling relational external functions is to
allow the environment to give arbitrary, not necessarily Boolean,
replies to the queries resulting from these symbols.  Then we could
have non-Boolean values for Boolean terms, and we would have to decide
how to handle this pathological situation, for example when it occurs
in the guard of a conditional rule.  In \cite[Section~5]{oa2}, this
approach was used, with the convention that this sort of pathology
would cause the conditional rule to fail.  In our present situation,
that convention no longer looks so natural, because the pathological
value might be one that the algorithm didn't really need.  (Recall
that in \cite{oa1,oa2,oa3} algorithms needed replies to all of their
queries.)  One can probably find a reasonable convention for dealing
with this pathology even for general interactive algorithms, but the
convention would appear somewhat arbitrary, and it seems simpler to
prohibit external function symbols from being relational.

It might appear that this prohibition could cause a problem in
programming.  Suppose, for example, that we know somehow that the
environment will provide a Boolean value for a certain nullary
external function symbol $p$.  Then we might want to use $p$ as the
guard in a conditional statement.  But we can't; since $p$ isn't a
relational symbol, it is not a Boolean term, and so (according to the
definitions in the following subsections) it is ineligible to serve as
a guard.  Fortunately, this problem disappears when we observe that
$p=\ttt{true}$ is a perfectly good guard (since equality is
relational) and it has the same value as $p$ (since we allegedly know
that $p$ gets a Boolean value).  If, on the other hand, we're not sure
that the environment will provide a Boolean value, then a particular
decision about how to handle a non-Boolean value can be built into the
program.  For example, the convention from \cite{oa2} would be given
by

  \begin{eatab}
     do in parallel\\
     \> if p = true then R1 endif\\
     \> if p = false then R2 endif\\
     \> if p $\ne$ true and p $\ne$ false then fail endif\\
     enddo.  
\end{eatab}

If one wanted to adopt a convention such as this, not only in a
particular program but throughout some programming language, then
one could adjoin external relational symbols to our syntax and treat
them as syntactic sugar for pieces of code like that exhibited above.  
\end{rmk}

\subsection{Terms}

\begin{df}
  The set of \emph{terms} is the smallest set containing
  $f(t_1,\dots,t_n)$ whenever it contains $t_1,\dots,t_n$ and $f$ is
  an $n$-ary function symbol from $\U\cup\E$.  (The basis of
  this recusive definition is, of course, given by the 0-ary function
  symbols.)
\qed\end{df}

This definition formalizes the assertion above that the external
function symbols in $\E$ are treated syntactically like those of the
state vocabulary $\U$.

Notice that the terms of ASMs do not involve variables.  In this
respect they differ from those of \cite{oa2}, those of first-order
logic, and those used in the bounded exploration witnesses of
\cite{ga1}. It may be surprising that we can get by without
variables while describing algorithms more general than those of
\cite{oa2} where we used variables.  Recall, however, that the
variables in the ASM programs of \cite{oa2} are bound by the
\ttt{let} construct, and that this construct is eliminable according
to \cite[Section~7]{oa3}.  In the present paper, we use \ttt{let}
only as syntactic sugar (see Subsection~\ref{sugar} below), and so
we do not need variables in our basic formalism.

\begin{df}
  A \emph{Boolean term} is a term of the form $f(\bld t)$ where $f$ is
  a relational symbol.
\qed\end{df}

\begin{conn}
  By $\Upsilon$-\emph{terms}, we mean terms built using the function
  symbols in $\Upsilon$ and variables.  These are terms in the usual
  sense of first-order logic for the vocabulary $\Upsilon$.  Terms as
  defined above, using function symbols from $\Upsilon\cup\E$ but not
  using variables, will be called \emph{ASM-terms} when we wish to
  emphasize the distinction from $\Upsilon$-terms.  A term of the form
  $f(\bld t)$ where $f \in \E$ will be called a \emph{query-term} or
  simply \emph{q-term}.
\end{conn}

The evaluation of an $\U$-term (in a given state with given values for
the variables) produces an element of the state, the value of the
term.  The same applies to q-terms, but there the situation is more
involved.  Consider a q-term $s = f(\bld t)$ and suppose that $\bld t$
has been evaluated to $\bld a$.  First the evaluation of $f(\bld a)$
produces a query, called the q-value of $s$.  If and when a reply to
the query is received the evaluation of $s$ is complete and we get the
actual value $s$.  See details in section~\ref{asm-sem}.

\subsection{Guards}

In \cite{oa2}, the guards $\phi$ in conditional rules \ttt{if} $\phi$
\ttt{then} $R_0$ \ttt{else} $R_1$ \ttt{endif} were simply Boolean
terms.  We shall need guards of a new sort to enable our ASMs to take
into account the temporal order of the environment's replies and to
complete a step even when some queries have not yet been answered.

We introduce timing explicitly into the formalism with the notation
$(s\preceq t)$, which is intended to mean that the replies needed to
evaluate the term $s$ arrived no later than those needed to evaluate
$t$.  It may seem that we are thereby just introducing a new
form of Boolean term, but in fact the situation is more complicated.

In the presence of
all the replies needed for both $s$ and $t$, the guard $s\preceq t$
will have a truth value, determined by relative timing of replies.
At the other extreme, if neither $s$ nor $t$ can be
fully evaluated,
then $(s\preceq t)$ must, like $s$ and
$t$ themselves, have no value.  So far, $(s\preceq t)$ behaves like
a term.

Between the two extremes, however, there are situations where
the replies provided by the environment suffice for the evaluation
of one but not both of $s$ and $t$. If replies suffice for $s$ but
not for $t$, then $(s\preceq t)$ is true; if replies suffice for $t$
but not for $s$, then $(s\preceq t)$ is false. Here, $(s\preceq t)$
behaves quite differently from a term, in that it has a value even
when one of its subterms does not.

This behavior of $(s\preceq t)$ also enables an ASM to complete its
step while some of its queries remain unanswered.  The execution of a
conditional rule with $(s\preceq t)$ as its guard can proceed to the
appropriate branch as soon as it has received enough replies from the
environment to evaluate at least one of $s$ and $t$, without waiting
for the replies needed to evaluate the other.

We shall need similar behavior for more complicated guards, and for
this purpose we shall use the propositional connectives of Kleene's
strong three-valued logic, which perfectly fits this sort of situation
\cite[\S64]{Kleene}.  We use the notations $\kand$ and $\kor$ for the
conjunction and disjunction of this logic.  They differ from the
classical connectives $\land$ and $\lor$ in that $\phi\kand\psi$ has
the value false as soon as either of $\phi$ and $\psi$ does, even if
the other has no value, and $\phi\kor\psi$ has the value true as soon
as either of $\phi$ and $\psi$ does, even if the other has no value.
In other words, if the truth value of one of the constituents $\phi$
and $\psi$ suffices to determine the truth value of the compound
formula, regardless of what truth value the other constituent gets,
then this determination takes effect without waiting for the other
constituent to get any truth value at all.  (It is customary, in
discussions of these modified connectives, to treat ``unknown'' as a
third truth value, but it will be convenient for us to regard it as
the absence of a truth value.  Such absences occur anyway, even for
ordinary terms, when the
existing replies do not suffice for a complete
evaluation, and it seems superfluous to introduce another entity,
``unknown,'' to serve as a marker of this situation.)

For detailed formal definition of the semantics of guards see
section \ref{asm-sem} below.

\begin{df}
  The set of \emph{guards} is defined by the following recursion.
  \begin{ls}
    \item Every Boolean term is a guard.
    \item If $s$ and $t$ are terms, then $(s\preceq t)$ is a guard.
    \item If $\phi$ and $\psi$ are guards, then so are
    $(\phi\kand\psi)$, $(\phi\kor\psi)$, and $\neg\phi$.
  \end{ls}
\qed\end{df}

Notice that the first clause of this definition allows, in particular,
terms built by means of the ordinary, 2-valued connectives from other
Boolean terms.

\subsection{Rules}

Most of the definition of ASM rules is as in \cite{oa2}.  The
differences are in the use of \ttt{issue} rules in place of the
less general \ttt{Output} rules of \cite{oa2} and in the more general
notion of guard introduced above.

\begin{df}
The set of ASM \emph{rules} is defined by the following recursion.
\begin{ls}
\item If $f\in\U$ is a dynamic $n$-ary function symbol, if
  $t_1,\dots,t_n$ are terms, and if $t_0$ is a term that is Boolean if
  $f$ is relational, then
$$
f(t_1,\dots,t_n):=t_0
$$
is a rule, called an \emph{update} rule.
\item If $f\in \E$ is an external $n$-ary function symbol and if
  $t_1,\dots,t_n$ are terms, then
$$
\ttt{issue}\ f(t_1,\dots,t_n)
$$
is a rule, called an \emph{issue} rule.
\item \ttt{fail} is a rule.
\item If $\phi$ is a guard and if $R_0$ and $R_1$ are rules, then
$$
\ttt{if\ } \phi\
\ttt{\ then\ } R_0 \ttt{\ else\ } R_1 \ttt{\ endif}
$$
is a rule, called a \emph{conditional} rule.  $R_0$ and $R_1$ are its
\emph{true} and \emph{false branches}, respectively.
\item If $k$ is a natural number (possibly zero) and if
  $R_1,\dots,R_k$ are rules then
$$
\ttt{do\ in\ parallel\ }R_1,\dots,R_k\ttt{\ enddo}
$$
is a rule, called a \emph{parallel combination} or \emph{block} with
the subrules $R_i$ as its \emph{components}.
\end{ls}
\qed\end{df}

We may omit the end-markers \ttt{endif} and \ttt{enddo} when they are
not needed, for example in very short rules or in programs formatted
so that indentation makes the grouping clear.

\begin{ex} \label{ex:broker} In \cite{ga1} we have analyzed the
algorithm of a broker who offers a block of $a$ shares of stock $s$
at price $p$ to clients $i$ by issuing queries $q_i(s,p,a)$,
$i=0,1$. The client whose reply reaches the broker first wins the
sale. We consider here a variant of the example in which every reply
from a client is considered to be positive, so that a client refuses
the offer by not answering at all. If both replies reach the broker
simultaneously then, for simplicity, client 0 is preferred. There is
a further timeout query $t$, so that if no client replies by
timeout, the sale is canceled. Given that $t, q_i \in \E$
and $s,p,a,0,1$ are some $\U$-terms, an equivalent ASM program
might be

\begin{eatab}
  \>\>\ttt{if\ }$\neg(q_0(s,p,a) \preceq t) \kand\neg( q_1(s,p,a)
  \preceq t)$
  \ttt{\ then\ cancel}\\
  \>\>\ttt{else\ if\ }$q_0(s,p,a) \preceq q_1(s,p,a)$\ttt{\ then\ sell\ to\ }0\\
  \>\>\ttt{else\ sell\ to\ }1
\end{eatab}

\noindent where \texttt{cancel} and \texttt{sell to }$i$ stand for
some updates recording respectively canceling the sale or selling to
client $i$ in the state.
\end{ex}

\subsection{Queries and templates}

We recall the query-reply model that is discussed at length in
\cite{oa1,oa2} and summarized in \cite{ga1}.  In addition to vocabulary
$\U$ and external vocabulary $\E$, an ASM has a set $\Lambda$ of
\emph{labels}.

\begin{df}
A \emph{potential query} in $\U$-structure $X$ is a finite
tuple of elements of $X\sqcup\Lambda$.  A \emph{potential reply} in
$X$ is an element of $X$. \qed\end{df}

Here $X\sqcup\Lambda$ is the disjoint union of $X$ and $\Lambda$.
So if they are not disjoint, then they are to be replaced by
disjoint isomorphic copies.  We shall usually not mention these
isomorphisms; that is, we write as though $X$ and $\Lambda$ were
disjoint.

The correspondence between external function calls on the one hand
and queries on the other hand is mediated by a template assignment,
defined as follows.

\begin{df}
For a fixed label set $\Lambda$, a \emph{template} for $n$-ary
  function symbols is any tuple in which certain positions are filled
  with labels from $\Lambda$ while the rest are filled with the
  \emph{placeholders} $\#1,\dots,\#n$, occurring once
  each.  We assume that these placeholders are distinct from all
  the other symbols under discussion ($\U\cup \E \cup\Lambda$).
  If $Q$ is a template for $n$-ary functions, then we write
  $Q[a_1,\dots,a_n]$ for the result of replacing each placeholder
  $\#i$ in $Q$ by the corresponding $a_i$.
  \qed\end{df}

Thus if the $a_i$ are elements of a state $X$ then
$Q[a_1,\dots,a_n]$ is a potential query in $X$.

\begin{df}
For a fixed label set and external vocabulary, a \emph{template
assignment} is a function assigning to each $n$-ary external
function symbol $f$ a template $\hat f$ for $n$-ary functions.
\qed\end{df}

The intention, which will be formalized in the semantic definitions
of the next section, is that when an ASM evaluates a term
$f(t_1,\dots,t_n)$ where $f\in \E$, it first computes the
values $a_i$ of the terms $t_i$, then issues the query $\hat
f[a_1,\dots,a_n]$, and finally uses the answer to this query as the
value of $f(t_1,\dots,t_n)$.

Template assignments solve the problem whether two distinct syntactic
occurrences of the same function symbol with the same arguments refer
to the same query or denote distinct queries.  Sometimes it is
convenient to have it one way, and sometimes another.  For extensive
discussion of template assignments we refer the reader to \cite{oa2}.

\subsection{Programs} \label{sugar}

Now we are ready to define ASM programs.

\begin{df}   \label{asm-prog-def}
  An \emph{interactive, small-step, ASM program} $\Pi$ consists of
  \begin{ls}
    \item a finite vocabulary $\U$,
    \item a finite set $\Lambda$ of labels,
    \item a finite external vocabulary $\E$,
    \item a rule $R$, using the vocabularies $\U$ and
    $\E$, the \emph{underlying rule} of $\Pi$,
    \item a template assignment with respect to $\E$ and $\Lambda$.
  \end{ls}
\end{df}

This completes the definition of the syntax of ASMs.  It will, however, be
convenient notationally and suggestive conceptually to introduce
abbreviations, syntactic sugar, for certain expressions.  Specifically, we
adopt the following conventions and notations.


\begin{conn}
  We use \ttt{skip} for the parallel combination with no components,
  officially written \ttt{do\ in\ parallel\ enddo}.
\end{conn}

\begin{conn}
The parallel combination with $k\geq2$ components $R_1,\ldots,R_k$ can
     be written as $R_1\ttt{\ par\ }\dots\ttt{\ par\ }R_k$.
\end{conn}

Semantically, \ttt{par} is commutative and associative, that is, rules
that differ only by the order and parenthesization of parallel
combinations will have the same semantic behavior.  Thus, in contexts
where only the semantics matters, parentheses can be omitted in
iterated \ttt{par}s.

\begin{conn}
  We abbreviate \ttt{if} $\phi$ \ttt{then} $R$ \ttt{else\ skip\ endif}
  as \ttt{if} $\phi$ \ttt{then} $R$ \ttt{endif}.
\qed\end{conn}

\begin{conn}
  For any term $t$, the Boolean term $t=t$ is denoted by $t!$, read as
  ``$t$ bang.''
\qed\end{conn}

These bang terms may seem trivial, but they can be used to control
timing in the execution of an ASM.  If the term $t$ involves external
function symbols, then the rule \ttt{if} $t!$ \ttt{then} $R$
\ttt{endif} differs from $R$ in that it issues the queries needed for
the evaluation of $t$ and waits for the replies before proceeding to
execute $R$.

\begin{conn}
  We use the following abbreviations:
\begin{align*}
    &(s \prec t)   &&\text{for} &&\neg(t \preceq s),  \\
    &(s \approx t)   &&\text{for} &&(s \preceq t) \kand (t \preceq s),\\
    &(s \succeq t) &&\text{for} &&(t \preceq s), \text{and}\\
    &(s \succ t)   &&\text{for} &&(t \prec s)
\end{align*}
Parentheses may be omitted when no confusion results.
\qed\end{conn}

The final two items of syntactic sugar involve two ways of binding
variables to terms by \ttt{let} operators.  Our syntax so far does not
include variables, but it is easy to add them.

\begin{df}
  Fix an infinite set of variables.  ASM \emph{rules with variables}
  are defined exactly like ASM rules, with variables playing the role
  of additional, nullary, static symbols.
\qed\end{df}

\begin{conn}
  If $R(v_1,\dots,v_k)$ is a rule with distinct variables $v_i$, and
  if $t_1,\dots,t_k$ are terms then the \emph{let-by-name}
  notation
$$
\nlet v_1=t_1,\dots,v_k=t_k\ttt{\ in\ }R(v_1,\dots,v_k)
$$ means $R(t_1,\dots,t_k)$.  \qed\end{conn}

\begin{conn}
  If $R(v_1,\dots,v_k)$ is a rule with distinct variables $v_i$, and
  if $t_1,\dots,t_k$ are terms then the \emph{let-by-value}
  notation
$$
\vlet v_1=t_1,\dots,v_k=t_k\ttt{\ in\ }R(v_1,\dots,v_k)
$$
abbreviates
$$
\ttt{if\ }t_1!\land\dots\land t_k!\ttt{\ then\ }R(t_1,\dots,t_k).
$$
\qed\end{conn}

For both \nlet and \vlet rules, the $v_i$ are called the
\emph{variables} of the rule, the $t_i$ its \emph{bindings}, and
$R(v_1,\dots,v_k)$ its \emph{body}.  Each of the variables $v_i$ is
bound by this rule at its initial occurrence in the context $v_i=t_i$
and at any free occurrences in $R(v_1,\dots,v_k)$.  (Occurrences of
the variables $v_i$ in the terms $t_j$ are not bound by the \nlet or
\vlet construction, regardless of whether $i=j$ or not.)

The let-by-name notation simply uses variables $v_i$ as placeholders for
the terms $t_i$.  The let-by-value notation, in contrast, first evaluates
all the $t_i$ and only afterward proceeds to execute the rule $R$.  In
this sense, the two forms of \ttt{let} correspond to call-by-name and
call-by-value in other situations.

\begin{ex} \label{ex:let} Let $t$ be a term representing a query asking the
environment for a fresh object, like constructors in object-oriented
languages, so that distinct textual occurrences of $t$ in a program
represent distinct queries with supposedly distinct replies.  Let $R(t)$
be a rule with several syntactic occurrences of $t$.  Then $\nlet x=t
\ttt{\ in\ }R(x)$ provides just an abbreviation for $R(t)$ (if it is
indeed shorter than $R(t)$), while $\vlet x=t \ttt{\ in\ }R(x)$ has a
completely different meaning: first ask the environment for a fresh
object, await the reply, and then use it repeatedly.
\end{ex}

\section{Interactive Small-Step ASMs: Semantics}   \label{asm-sem}

Throughout this section, we refer to a fixed structure $X$.  We start by
recalling the notion of history introduced and motivated in the companion
paper \cite{ga1}.  Then we define the semantics of terms, guards, and
rules in the structure $X$, relative to histories $\xi$.  In each case, we
tacitly presume a template assignment.  (Unlike $X$, the history $\xi$
will not remain fixed, because the meaning of a guard under history $\xi$
can depend on the meanings of its subterms under initial segments of
$\xi$.)  In each case, the semantics will specify a causality relation.
In addition, for terms and guards the semantics may provide a value
(Boolean in the case of guards); for rules, the semantics may declare the
history final, successful, or failing, and may provide updates.

\subsection{Histories}

The notion of \emph{history} as a formal model of intrastep interaction of
an algorithm and its environment has been introduced and extensively
discussed in \cite{ga1}.  We recall the relevant definitions.

\begin{df}       \label{ans-fn}
An \emph{answer function} for a state $X$ is a partial map from
potential queries to potential replies.  A \emph{history} for $X$ is
a pair $\xi=\sq{\ans\xi,\leq_\xi}$ consisting of an answer function
$\ans\xi$ together with a linear pre-order $\leq_\xi$ of its domain.
By the \emph{domain} of a history $\xi$, we mean the domain
\dom{\ans\xi} of its answer function component, which is also the
field of its pre-order component. \qed\end{df}

Recall that a \emph{pre-order} of a set $D$ is a reflexive,
transitive, binary relation on $D$, and that it is said to be
\emph{linear} if, for all $x,y\in D$, $x\leq y$ or $y\leq x$.  The
equivalence relation defined by a pre-order is given by
$$
x\equiv y\iff x\leq y\leq x.
$$
The equivalence classes are partially ordered by
$$
[x]\leq[y]\iff x\leq y,
$$
and this partial order is linear if and only if the pre-order was.

The \emph{length} of a linear pre-order is defined to be the order
type of the induced linear ordering of equivalence classes.  (We
shall use this notion of length only in the case where the number of
equivalence classes is finite, in which case this number serves as
the length.)

We also write $x<y$ to mean $x\leq y$ and $y\not\leq x$.  (Because a
pre-order need not be antisymmetric, $x<y$ is in general a stronger
statement than the conjunction of $x\leq y$ and $x\neq y$.)  When,
as in the definition above, a pre-order is written as $\leq_\xi$, we
write the corresponding equivalence relation and strict order as
$\equiv_\xi$ and $<_\xi$.  The same applies to other subscripts and
superscripts.

\begin{df}
Let $\leq$ be a pre-order of a set $D$.  An \emph{initial segment}
of $D$ with respect to $\leq$ is a subset $S$ of $D$ such that
whenever $x\leq y$ and $y\in S$ then $x\in S$.  An \emph{initial
segment} of $\leq$ is the restriction of $\leq$ to an initial
segment of $D$ with respect to $\leq$.  An \emph{initial segment} of
a history \sq{{\ans\xi},\leq_\xi} is a history \sq{{\ans\xi}\restr
S,\leq_\xi\restr S}, where $S$ is an initial segment of
\dom{\ans\xi}\ with respect to $\leq_\xi$.  (We use the standard
notation $\restr$ for the restriction of a function or a relation to
a set.)  We write $\eta\initeq\xi$ to mean that the history $\eta$
is an initial segment of the history $\xi$. \qed\end{df}

\subsection{Terms}

The semantics of terms presumes not only an $\U$-structure $X$ and a
template assignment but also a history $\xi$.  The semantics is
essentially the same as in \cite{oa2}, except that we do not use variables
here.  In particular, the history $\xi$ is involved only via the answer
function $\ans\xi$; the pre-order is irrelevant.

The semantics of terms specifies, by induction on terms $t$, the queries
that are caused by $\xi$ under the associated causality relation
$\vdash^t_X$ and sometimes also a value $\val tX\xi$.  In the case of
query-terms, the semantics may specify also a query-value $\qval tX\xi$.
An evaluation of a query-term $t$ is intended to produce first a query,
called the q-value of $t$ and denoted $\qval tX\xi$; the reply, if any, to
the query is the actual value $\val tX\xi$ of $t$.

\begin{df}   \label{term-val-def}
  Let $t$ be the term $f(t_1,\dots,t_n)$.
  \begin{ls}

   \item If \val{t_i}X\xi\ is undefined for at least one $i$, then
    \val tX\xi\ is also undefined, and $\xi\vdash^t_Xq$ if and only
    if $\xi\vdash^{t_i}_Xq$ for at least one $i$. If $f \in
    \E$ then $\qval tX\xi$ is also undefined.

   \item If, for each $i$, $\val{t_i}X\xi=a_i$ and if $f\in\U$,
    then $\val tX\xi=f_X(a_1,\dots,a_n)$, and no query $q$ is
    caused by $\xi$.

   \item If, for each $i$, $\val{t_i}X\xi=a_i$, and if $f\in \E$, then
    $\qval tX\xi$ is the query $\hat f[a_1,\dots,a_n]$.

   \begin{ls}

    \item If $\qval tX\xi = q \in \dom{\ans \xi}$, then $\val
    tX\xi=\ans\xi(q)$, and no query is caused by $\xi$.

    \item If $\qval tX\xi = q \notin \dom{\ans \xi}$, then $\val tX\xi$ is
    undefined, and $q$ is the unique query such that $\xi\vdash^t_Xq$.
    \end{ls}
  \end{ls}
\qed\end{df}

We record for future reference three immediate consequences of this
definition; the proofs are routine inductions on terms.

\begin{la}   \label{df-noq-t}
  \val tX\xi\ is defined if and only if there is no query
$q$ such that $\xi\vdash^t_Xq$.
\end{la}

\begin{la}   \label{no-rep-t}
If $\xi\vdash^t_Xq$ then $q\notin\dom{\ans\xi}$.
\end{la}

\begin{la} \label{mon-t} If $\eta\initeq\xi$ (or even if merely
  $\ans\eta\subseteq\ans\xi$) and if \val tX\eta\ is defined, then
  \val tX\xi\ is also defined and these values are equal.  Similarly,
  if $t$ is a q-term such that \qval tX\eta\ exists, then $\qval
  tX\eta=\qval tX\xi$.

\end{la}

\subsection{Guards}

The semantics of guards, unlike that of terms, depends not only on the
answer function but also on the preorder in the history.  Another
difference from the term case is that the values of guards, when
defined, are always Boolean values.  Guards share with terms the
property that they produce queries if and only if their values are
undefined.

\begin{df}   \label{guard-sem-def}
Let $\phi$ be a guard and $\xi$ a history in an $\U$-structure
$X$.
\begin{ls}
  \item If $\phi$ is a Boolean term, then its value (if any) and
  causality relation are already given by
  Definition~\ref{term-val-def}.
  \item If $\phi$ is $(s\preceq t)$ and if both $s$ and $t$ have values
  with respect to $\xi$, then $\val\phi X\xi=\ttt{true}$ if, for every
  initial segment $\eta\initeq\xi$ such that \val tX\eta\ is defined,
  $\val sX\eta$ is also defined.  Otherwise, $\val\phi
  X\xi=\ttt{false}$.  Also declare that $\xi\vdash^\phi_Xq$ for no
  $q$.
  \item If $\phi$ is $(s\preceq t)$ and if $s$ has a value with respect
  to $\xi$ but $t$ does not, then define $\val\phi X\xi$ to be
  \ttt{true}; again declare that $\xi\vdash^\phi_Xq$ for no $q$.
  \item If $\phi$ is $(s\preceq t)$ and if $t$ has a value with respect
  to $\xi$ but $s$ does not, then define $\val\phi X\xi$ to be
  \ttt{false}; again declare that $\xi\vdash^\phi_Xq$ for no $q$.
  \item If $\phi$ is $(s\preceq t)$ and if neither $s$ nor $t$ has a
  value with respect to $\xi$, then $\val\phi X\xi$ is undefined,
  and $\xi\vdash^\phi_Xq$ if and only if $\xi\vdash^s_Xq$ or
  $\xi\vdash^t_Xq$.
  \item If $\phi$ is $\psi_0\kand\psi_1$ and both $\psi_i$ have value
  \ttt{true}, then $\val\phi X\xi=\ttt{true}$ and no query is
  produced.
  \item If $\phi$ is $\psi_0\kand\psi_1$ and at least one $\psi_i$ has
  value \ttt{false}, then $\val\phi X\xi=\ttt{false}$ and no query
  is produced.
  \item If $\phi$ is $\psi_0\kand\psi_1$ and one $\psi_i$ has value
  \ttt{true} while the other, $\psi_{1-i}$, has no value, then
  $\val\phi X\xi$ is undefined, and $\xi\vdash^\phi_Xq$ if and only if
  $\xi\vdash^{\psi_{1-i}}_Xq$.
  \item If $\phi$ is $\psi_0\kand\psi_1$ and neither $\psi_i$ has a
  value, then \val\phi X\xi\ is undefined, and $\xi\vdash^\phi_Xq$
  if and only if $\xi\vdash^{\psi_i}_Xq$ for some $i$.
  \item The preceding four clauses apply with $\kor$ in place of
  $\kand$ and \ttt{true} and \ttt{false} interchanged.
  \item If $\phi$ is $\neg\psi$ and $\psi$ has a value, then
  $\val\phi X\xi=\neg\val\psi X\xi$ and no query is produced.
  \item If $\phi$ is $\neg\psi$ and $\psi$ has no value then
  \val\phi X\xi\ is undefined and $\xi\vdash^\phi_Xq$ if and only
  if $\xi\vdash^{\psi}_Xq$.
\end{ls}
\qed\end{df}

\begin{rmk}
An alternative, and perhaps more intuitive, formulation of the
definition of $\val{(s\preceq t)}X\xi$ in the case where both $s$ and
$t$ have values is to let $\xi'$ (resp.\ $\xi''$) be the shortest
initial segment of $\xi$ with respect to which $s$ (resp.\ $t$) has a
value, and to define $\val\phi X\xi$ to be \ttt{true} if
$\xi'\initeq\xi''$ and \ttt{false} otherwise.  This is equivalent, in
the light of Lemma~\ref{mon-t}, to the definition given above, but it
requires knowing that the shortest initial segments mentioned here,
$\xi'$ and $\xi''$, exist.  That is clearly the case if the partial
order associated to the preorder in $\xi$ is a well-ordering, in
particular if it is finite.  Once we establish that ASMs satisfy the
Bounded Work Postulate, it will follow that we can confine our
attention to finite histories and so use the alternative explanation
of \val{(s\preceq t)}X\xi.  The formulation adopted in the definition
has the advantage of not presupposing that only finite histories
matter.
\qed\end{rmk}

\begin{ex}
The truth value of a timing guard $(s\preceq t)$ is defined in terms of
the syntactic objects $s$ and $t$, not in terms of their values.  As a
result, this truth value may not be preserved if $s$ and $t$ are replaced
by other terms with the same values (in the given history $\xi$), not even
if the replacement terms ultimately issue the same queries as the original
ones.  Here is an example of what can happen.  Suppose $p$, $q$, and $r$
are external function symbols, $p$ being unary and the other two nullary.
Suppose further that 0 is a static nullary $\U$-symbol.  Consider a history
$\xi$ with three queries in its domain, pre-ordered as $\hat
p[0_X]<_\xi\hat q<_\xi\hat r$, and with $\ans\xi(\hat r)=0_X$.  Then the
term $p(0)$ has a value already for the initial segment of $\xi$ of length
1; $q$ gets a value later, namely for the initial segment of length 2; and
$p(r)$ gets a value only for the whole history $\xi$, of length 3.  Thus,
the guards $(p(0)\prec q)$ and $(q\prec p(r))$ are true, even though
$p(0)$ and $p(r)$ have the same value and have, as the ultimate step in
their evaluation, the answer to the query $\hat p[0_X]$.  \qed\end{ex}

Just as for terms, the following lemmas follow immediately, by
induction on guards, from the definition plus the corresponding lemmas
for terms.

\begin{la}   \label{df-noq-g}
  \val\phi X\xi\ is defined if and only if there is no query
$q$ such that $\xi\vdash^\phi_Xq$.
\end{la}

\begin{la}   \label{no-rep-g}
If $\xi\vdash^\phi_Xq$ then $q\notin\dom{\ans\xi}$.
\end{la}

\begin{la}   \label{mon-g}
If $\eta\initeq\xi$ and if \val\phi X\eta\ is defined, then \val\phi
  X\xi\ is also defined and these values are equal.
\end{la}

\begin{rmk}
  Given the semantics of guards, we can amplify the statement, in the
  Remark~3.10. of \cite{ga1}, that guards express descriptions like
  ``$p$ has reply $a$ and $p'$ has no reply.''  In view of
  Lemma~\ref{mon-g}, it is more accurate to say that a guard expresses
  that such a description either is correct now or was so at some
  earlier time.  The lemma says that, once a guard is true, it remains
  true when the history is extended by adding new elements later in
  the preorder, whereas a property like ``$p'$ has no reply'' need not
  remain true.  Thus, what a guard can really express is something
  like this: it either is now true or was once true that ``$p$ has
  reply $a$ and $p'$ has no reply yet.''  This particular example
  would be expressed by the guard $(p=a)\kand(p\prec p')$, where, for
  simplicity we have not introduced a separate notation for 0-ary
  symbols corresponding to the queries $p$ and $p'$ and the element
  $a$.
\end{rmk}

\subsection{Rules}

The semantics of a rule, for an $\U$-structure $X$, an appropriate
template assignment, and a history $\xi$, consists of a \emph{causality
relation}, declarations of whether $\xi$ is \emph{final} and whether it
\emph{succeeds} or \emph{fails}, and a set of \emph{updates}.

\begin{df}
  Let $R$ be a rule and $\xi$ a history for the $\U$-structure $X$.  In the
  following clauses, whenever we say that a history succeeds or that
  it fails, we implicitly also declare it to be final;
  contrapositively, when we say that a history is not final, we
  implicitly also assert that it neither succeeds nor fails.
  \begin{ls}
    \item If $R$ is an update rule $f(t_1,\dots,t_n):=t_0$ and if all
    the $t_i$ have values $\val{t_i}X\xi=a_i$, then $\xi$ succeeds for
    $R$, and it produces the update set
    $\{\sq{f,\sq{a_1,\dots,a_n},a_0}\}$ and no queries.
    \item If $R$ is an update rule $f(t_1,\dots,t_n):=t_0$ and if some
    $t_i$ has no value, then $\xi$ is not final for $R$, it produces
    the empty update set, and $\xi\vdash^R_Xq$ if and only if
    $\xi\vdash^{t_i}_Xq$ for some $i$.
  \item If $R$ is $\ttt{issue\,}f(t_1,\dots,t_n)$ and if all the $t_i$
  have values $\val{t_i}X\xi=a_i$, then $\xi$ succeeds for $R$, it
  produces the empty update set, and $\xi\vdash^R_Xq$ for the single
  query $q=\hat f[a_1,\dots,a_n]$ provided $q\notin\dom{\ans\xi}$; if
  $q\in\dom{\ans\xi}$ then no query is produced.
    \item If $R$ is $\ttt{issue\,}f(t_1,\dots,t_n)$ and if some $t_i$
    has no value, then $\xi$ is not final for $R$, it produces the
    empty update set, and $\xi\vdash^R_Xq$ if and only if
    $\xi\vdash^{t_i}_Xq$ for some $i$.
  \item If $R$ is \ttt{fail}, then $\xi$ fails for
    $R$; it produces the empty update set and no queries.
    \item If $R$ is a conditional rule $\ttt{if\ } \phi \ttt{\ then\ }
R_0 \ttt{\ else\ } R_1 \ttt{\ endif}$ and if $\phi$ has no value, then
$\xi$ is not final for $R$, and it produces the empty update set.
$\xi\vdash^R_Xq$ if and only if $\xi\vdash^\phi_Xq$.
     \item If $R$ is a conditional rule $\ttt{if\ } \phi
\ttt{\ then\ } R_0 \ttt{\ else\ } R_1 \ttt{\ endif}$ and if $\phi$ has
value \ttt{true} (resp.\ \ttt{false}), then finality, success, failure,
updates, and queries are the same for $R$ as for $R_0$ (resp.\ $R_1$).
     \item If $R$ is a parallel combination $\ttt{do\ in\ parallel\
     }R_1,\dots,R_k\ttt{\ enddo}$ then:
       \begin{ls}
     \item $\xi\vdash^R_Xq$ if and only if $\xi\vdash^{R_i}_Xq$
     for some $i$.
     \item The update set for $R$ is the union of the update sets
     for all the components $R_i$.  If this set contains two
     distinct updates at the same location, then we say that a
     \emph {clash} occurs (for $R$, $X$, and $\xi$).
     \item $\xi$ is final for $R$ if and only if it is final for
     all the $R_i$.
     \item $\xi$ succeeds for $R$ if and only if it succeeds for
         all the $R_i$ and no clash occurs.
     \item $\xi$ fails for $R$ if and only if it is final for $R$
     and either it fails for some $R_i$ or a clash occurs.
       \end{ls}
  \end{ls}
\qed\end{df}

There is no analog for rules of Lemmas~\ref{df-noq-t} and
\ref{df-noq-g}.  A rule may issue queries even though it is final
(in the case of an \ttt{issue} rule) or produces updates (in the
case of parallel combinations) or both.  There are, however, analogs
for the other two lemmas that we established for terms and guards;
again the proofs are routine inductions.

\begin{la}   \label{no-rep-r}
If $\xi\vdash^R_Xq$ then $q\notin\dom{\ans\xi}$.
\end{la}

\begin{la}   \label{mon-r}
Let $\eta\initeq\xi$.
\begin{ls}
  \item If $\eta$ is final for $R$, then so is $\xi$.
  \item If $\eta$ succeeds for $R$, then so does $\xi$.
  \item If $\eta$ fails for $R$, then so does $\xi$.
  \item The update set for $R$ under $\xi$ includes that under
  $\eta$.
\end{ls}
\end{la}

The reader might find it useful at this point to work out the
semantic details of the examples \ref{ex:broker} and \ref{ex:let},
comparing the results with the intuitive explanations given in the
examples.

\begin{rmk}
  Issue rules are the only way an ASM can issue a query without
  necessarily waiting for an answer.  More precisely, if a history
  causes a rule to issue a query and is also final for that rule, then
  that rule either is an issue rule or contains a subrule with the
  same property.  Thus, we cannot eliminate \ttt{issue} from the
  syntax without reducing the power of ASMs.
\end{rmk}

\subsection{ASM definition}

If $\xi$ is a successful, final history for a rule $R$ over an
$\U$-structure $X$, then $R$ and $\xi$ produce a successor for $X$.
We need a preliminary lemma to ensure that this successor will be
well-defined.  Recall that, in the definition of the semantics of
parallel rules, we defined ``a clash occurs'' (for a rule, template
assignment, state, and history) to mean that the update set contains
two different updates of the same location.

\begin{la}   \label{no-clash}
  If an ASM rule with a template assignment is (final and) successful
  in a certain state with a certain history, then no clash occurs for
  this rule, template assignment, state, and history.  
\end{la}

\begin{proof}
Use induction on rules.  In the case of a parallel composition, the
semantics explicitly provided for failure if a clash occurs.  All
other cases are trivial thanks to the induction hypothesis.
\end{proof}

\begin{df} \label{asm-next_state}
Fix a rule $R$ endowed with a template assignment, and let $X$ be an
$\U$-structure and $\xi$ be a history for $X$.  If $\xi$ is successful and
final for $R$ over $X$, and if $\DD(X,\xi)$ is the update set produced by
$R$, $X$, and $\xi$, then the \emph{successor} $\tau(X,\xi)$ of $X$ with
respect to $R$ and $\xi$ is the $\U$-structure $Y$ such that
\begin{ls}
  \item $Y$ has the same base set as $X$,
  \item $f_Y(\bld a)=b$ if $\sq{f,\bld a,b}\in\DD(X,\xi)$, and
  \item otherwise $Y$ interprets function symbols exactly as $X$ does.\qed
\end{ls}
\end{df}

Lemma~\ref{no-clash} ensures that the second clause of
the definition does not attempt to give $f_Y(\bld a)$ two different
values. 

Now we are ready to give a complete definition of ASMs.

\begin{df}   \label{asm-def}
  An \emph{interactive, small-step, ASM} consists of
  \begin{ls}
    \item an ASM program $\Pi$ in some vocabulary $\U$,
    \item a nonempty set $\scr S$ of $\U$-structures called \emph{states}
    of the ASM, and
    \item a nonempty set $\scr I\subseteq \scr S$ of \emph{initial states},
  \end{ls}
subject to the requirements that \scr S and \scr I are closed under
isomorphism and that \scr S is closed under transitions in the following
sense.  If $X\in\scr S$, if $\xi$ is a successful, final history for $\Pi$
in $X$, and if $\DD(X,\xi)$ is the update set produced by $\Pi$, $X$, and
$\xi$, then the successor $\tau(X,\xi)$ of $X$ with respect to $\Pi$ and
$\xi$ is also in \scr S.  The successor is the \emph{next state} for $X$
with respect to $\Pi$, endowed with the given template assignment, and to
$\xi$.  \qed
\end{df}

\section{ASMs are Algorithms} \label{sec:asms-are-algos}

This section is devoted to checking that ASMs, as just defined, are
algorithms, as defined in \cite{ga1}. In this section (and in the
rest of the paper) we freely use the notions and results of
\cite{ga1}.

\subsection{Obvious postulates}

Much of this checking is trivial: Everything required by the States
Postulate of \cite{ga1} is in our definition of ASMs.  The causality
relation required by the Interaction Postulate of \cite{ga1} is
included in our semantics for ASMs. (Strictly speaking, the
causality relation defined for ASMs should be restricted to finite
histories, to comply with the statement of the Interaction
Postulate.)  The Isomorphism Postulate of \cite{ga1} is also
obvious, because everything involved in our ASM semantics is
invariant under isomorphisms.  So the only postulates requiring any
real checking are the Step and Bounded Work Postulates.

\subsection{Step Postulate}

The ASM semantics provides notions of finality, success, failure, and
updates.  In addition to these, the Step Postulate of \cite{ga1}
requires (in Part~C) a notion of next state and (in Part~A) assurance
that every complete, coherent history has a final initial
segment\footnote{We implicitly use the notions of coherent history and
  complete history as defined for algorithms in general in
  \cite[Section~3]{ga1}, with respect to the causality relation of the
  ASM program as defined in Section~\ref{asm-sem} above}.  The next
state is given by Definition~\ref{asm-next_state}, and it is
well-defined because of Lemma~\ref{no-clash}.

To show that every complete, coherent history has a final initial
segment, we actually show more, namely that every complete history is
final.  The main ingredient here is the following lemma.

\begin{la}
  If a history $\xi$ is not final for a rule $R$ in a state $X$, then
  $\xi\vdash^R_Xq$ for some query $q$.
\end{la}

\begin{proof}
  We proceed by induction on the rule $R$, according to the clauses in
  the definition of the semantics for rules.  Since we are given that
  $\xi$ is not final, we can ignore those clauses that say $\xi$ is
  final, and there remain the following cases.

If $R$ is either an update rule $f(t_1,\dots,t_n):=t_0$ or an issue
rule $\ttt{issue}\ f(t_1,\dots,t_n)$ and some $t_i$ has no value, then
by Lemma~\ref{df-noq-t} there is a query $q$ such that
$\xi\vdash^{t_i}_Xq$, and therefore $\xi\vdash^R_Xq$.

If $R$ is a conditional rule whose guard has no value, then the same
argument applies except that we invoke Lemma~\ref{df-noq-g} in place
of Lemma~\ref{df-noq-t}.

If $R$ is a conditional rule whose guard has a truth value, then the
lemma for $R$ follows immediately from the lemma for the appropriate
branch of $R$.

Finally, suppose $R$ is a parallel combination.  Since $\xi$ is not
final for $R$ in $X$, there is a component $R_i$ for which $\xi$ is
not final.  By induction hypothesis, $\xi\vdash^{R_i}_Xq$ for some
$q$, and then we also have $\xi\vdash^R_Xq$.
\end{proof}

To complete the verification of the Step Postulate, we observe that,
in the situation of the lemma, $q\in\Issued^R_X(\xi)$ and, by
Lemma~\ref{no-rep-r}, $q\notin\dom{\ans\xi}$.  Thus,
$q\in\text{Pending}^R_X(\xi)$, and so $\xi$ is not complete for $R$
and $X$.

\begin{rmk}
  Because we have promised to prove that every algorithm is equivalent
  to an ASM, one might think that every algorithm enjoys the property
  established for ASMs in the preceding proof, namely that all
  complete histories are final.  This is, however, not the case,
  because this property is not preserved by equivalence of algorithms.
  For a simple example, consider an algorithm where, for every state,
  the empty history is the only final history, and it causes one
  query, while all other histories cause no queries.  Since the empty
  history is an initial segment of every history, Part~A of the Step
  Postulate is satisfied, even though the complete histories, those in
  which the one query is answered, are not final.

Notice, however, that converting an arbitrary algorithm to an
equivalent one in which all complete histories are final is much
easier than converting it to an equivalent ASM.  Simply adjoin all
non-final, complete histories for any state to the set of final,
failing histories for that state.  None of the histories newly
adjoined here can be attainable, so the modified algorithm is
equivalent to the original.  \qed\end{rmk}

\subsection{Bounded Work Postulate}
We turn now to the Bounded Work Postulate of \cite{ga1}.  Its first
assertion, about the lengths of queries, is easy to check.  Since
the postulate refers only to coherent histories (actually to
attainable, final histories, but coherence suffices for the present
purpose), any query in the domain of such a history is caused by
some history.  By inspection of the definition of ASM semantics, all
queries that are ever caused are of the form $\hat f[a_1,\dots,a_n]$
and thus have the same length as the template $\hat f$ assigned to
some external function symbol.  As there are only finitely many
external function symbols, the lengths of the queries are bounded.

The next assertion of the Bounded Work Postulate, bounding the number
of queries issued by the algorithm, will be a consequence of the
following lemma.

\begin{la}
For any term $t$, guard $\phi$, or rule $R$, there is a natural number
$B(t)$, $B(\phi)$, or $B(R)$ that bounds the number of queries caused
in a state $X$ by initial segments of a history $\xi$.  The bound
depends only on $t$, $\phi$, or $R$, not on $X$ or $\xi$.
\end{la}

\begin{proof}
Go to the definition of the semantics of ASMs and inspect the clauses
that say queries are caused.  The result is that, first, we can define
the desired $B(t)$ for terms by
$$
B(f(t_1,\dots,t_n))=1+\sum_{i=1}^nB(t_i).
$$
The sum here comes from the first clause in the definition of
semantics of terms, and the additional 1 comes from the last clause.
It is important here that, according to Lemma~\ref{mon-t}, all the
initial segments of any $\xi$ that produce values for a $t_i$ produce
the same value $a_i$.  Thus, the last clause of the definition
produces at most one query $\hat f[a_1,\dots,a_n]$.

Similarly, we obtain for guards $\phi$ (other than the Boolean terms
already treated above) the estimates
\begin{eqs}
B((s\preceq t))&=&B(s)+B(t)\\
B(\psi_0\kand\psi_1)=B(\psi_0\kor\psi_1)&=&B(\psi_0)+B(\psi_1)\\
B(\neg\psi)&=&B(\psi).
\end{eqs}

For rules, we obtain
\begin{eqs}
B(f(t_1,\dots,t_n):=t_0)&=&\sum_{i=0}^nB(t_i)\\
B(\ttt{issue}f(t_1,\dots,t_n))&=&1+\sum_{i=1}^nB(t_i)\\
B(\ttt{fail})&=&0\\
B(\ttt{if\ }\phi\ttt{\ then\ }R_0\ttt{\ else\ }R_1)&=&
B(\phi)+B(R_0)+B(R_1)\\
B(\ttt{do\ in\ parallel\ }R_1,\dots R_k)&=&
\sum_{i=1}^kB(R_i).
\end{eqs}%
(In the bound for conditional rules, we could reduce $B(R_0)+B(R_1)$
to $\max\{B(R_0),B(R_1)\}$ by using the fact that all the initial
segments of any $\xi$ that produce values for $\phi$ produce
the same value.)
\end{proof}

Since $\Issued^R_X(\xi)$ is the set of queries caused in state $X$,
under rule $R$, by initial segments of $\xi$, the lemma tells us that
$|\Issued^R_X(\xi)|\leq B(R)$, independently of $X$ and $\xi$.  This
verifies the second assertion of the Bounded Work Postulate.  (It
actually verifies more, since the proof applies to all histories
$\xi$, not merely to attainable ones.)

To complete the verification of the Bounded Work Postulate, it remains
only to produce bounded exploration witnesses for all ASMs.  We shall
do this by an induction on rules, preceded by proofs of the analogous
results for terms and for guards.

\begin{la}
For every ASM-term $t$ (without variables) there exists a finite set
$W(t)$ of $\U$-terms (possibly with variables) such that,
whenever $(X,\xi)$ and $(X',\xi)$ agree on $W(t)$, then:
\begin{ls}
  \item If $\xi\vdash^t_Xq$ then $\xi\vdash^t_{X'}q$.
  \item $\val tX\xi=\val t{X'}\xi$.
\end{ls}
\end{la}

Recall that ``agree on $W(t)$'' means that each term in $W(t)$ has the
same value in  $X$ and in $X'$ when the variables are given the same
values in \ran{\ans\xi}.  Recall also that an equation between
possibly undefined expressions like \val tX\xi\ means that if either
side is defined then so is the other and they are equal.

\begin{proof}
By a \emph{shadow} of an ASM-term $t$, we mean a term $\tilde t$
obtained from $t$ by putting distinct variables in place of the
outermost\footnote{``Outermost'' means ``maximal'' in the sense that
the occurrence in question is not properly contained in another such
occurrence.  In terms of the parse tree of $t$, it means that, on the
path from the root of the whole tree to the root of the subtree given
by the occurrence in question, there is no other occurrence of an
external function symbol.} occurrences of subterms that begin
with external function symbols.  Thus, $\tilde t$ is an
$\U$-term, and $t$ can be recovered from $\tilde t$ by a
suitable substitution of ASM-terms (that start with external function
symbols) for all the variables.

Notice that $\tilde t$ fails to be uniquely determined
by $t$ only because we have not specified which variables are to replace
the subterms.

We define, by recursion on ASM-terms $t$, the set $W(t)$ of
$\U$-terms as follows.  If $t$ is $f(t_1,\dots,t_n)$ then
$$
W(t)=\{\tilde t\}\cup\bigcup_{i=1}^nW(t_i),
$$
where $\tilde t$ is some shadow of $t$. It follows immediately, by
induction on $t$, that $W(t)$ is finite.  The verification that this
$W(t)$ satisfies the conclusion of the lemma is also by induction on
$t$, following the clauses in the definition of the semantics of
terms.

Assume that $(X,\xi)$ and $(X',\xi)$ agree on $W(t)$.  Notice that
they also agree on each $W(t_i)$, because $W(t_i)\subseteq W(t)$.

Suppose first that \val {t_i}X\xi\ is undefined for some $i$.  By
induction hypothesis, \val{t_i}{X'}\xi\ is also undefined, so the same
clause of the semantics of terms applies in $X$ and $X'$.  That clause
says that $t$ has no value in either state and it issues those queries
that are issued by any of the $t_i$.  Those are the same queries in
$X$ as in $X'$ by the induction hypothesis.

{}From now on, suppose that $\val{t_i}X\xi=a_i$ for each $i$.  By
induction hypothesis, the same holds for $X'$, with the same $a_i$'s.

So if $f\in\U$ then $t$ gets the value $f_X(a_1,\dots,a_n)$ in
$X$ and the value $f_{X'}(a_1,\dots,a_n)$ in $X'$, and we must check
that these values are the same.  Recall that $t$ is obtained from its
shadow $\tilde t$ by replacing each variable $v$ in $\tilde t$ with a
certain ASM-term $\sigma(v)$.  Thus, the value $f_X(a_1,\dots,a_n)$ of
$t$ in $X$ is also the value of $\tilde t$ in $X$ when each variable
$v$ is assigned the value $\val{\sigma(v)}X\xi$ and similarly with
$X'$ in place of $X$.  By induction hypothesis, these values assigned
to the variables are the same in $X$ and $X'$.  (We use here that
$\sigma(v)$ is a \emph{proper} subterm of $t$, which is correct
because $t$ begins with a function symbol from $\U$.)
Furthermore, since $\sigma(v)$ begins with an external function
symbol, its value is in \ran{\ans\xi}.  Thus, the assumption that
$(X,\xi)$ and $(X',\xi)$ agree on $W(t)$, which contains $\tilde t$,
ensures that $\tilde t$ has the same value in both $X$ and $X'$.
Therefore $f_X(a_1,\dots,a_n)=f_{X'}(a_1,\dots,a_n)$ as required.
Since no queries are issued in this situation, we have completed the
proof in the case that $f\in\U$.

There remains the case that $f\in\E$ and, as before, the subterms
$t_i$ have (the same) values $a_i$ in $X$ and $X'$.  If $\hat
f[a_1,\dots,a_n]\in\dom{\ans\xi}$ then $t$ gets the same value
$\ans\xi(\hat f[a_1,\dots,a_n])$ in both $X$ and $X'$, no queries are
issued in either state, and the lemma is established in this case.

So assume that the query $\hat f[a_1,\dots,a_n]$ is not in \dom{\ans\xi}.
Then this query is the unique query produced by $t$ in either state,
and $t$ has no value in either state, so again the conclusion of the
lemma holds.
\end{proof}

The preceding lemma easily implies the corresponding result for
guards.

\begin{la}
For every guard $\phi$ there exists a finite set $W(\phi)$ of
$\U$-terms such that, whenever $(X,\xi)$ and $(X',\xi)$ agree on
$W(\phi)$, then:
\begin{ls}
  \item If $\xi\vdash^\phi_Xq$ then $\xi\vdash^\phi_{X'}q$.
  \item $\val\phi X\xi=\val\phi{X'}\xi$.
\end{ls}
\end{la}

\begin{proof}
We define $W(\phi)$ by induction on $\phi$.  If $\phi$ is a Boolean
term, then the preceding lemma provides the required $W(\phi)$.

If $\phi$ is $(s\preceq t)$ then we define
$$
W(s\preceq t)=W(s)\cup W(t)\cup\{\ttt{true,false}\}.
$$
To check that the conclusion of the lemma is satisfied, we apply the
previous lemma to see that, not only for the history $\xi$ in question
but also for any $\eta\initeq\xi$, if either of \val sX\eta\ and \val
s{X'}\eta\ is defined then so is the other, and similarly for $t$.
With this information and
with the knowledge that \ttt{true} and \ttt{false} denote the same
element in $X$ and $X'$ (because of agreement on $W(\phi)$, which
contains \ttt{true} and \ttt{false}), one finds by inspection of the
relevant clauses in the semantics of guards that the conclusion of the
lemma holds.

If $\phi$ is $\psi_0\kand\psi_1$ or $\psi_0\kor\psi_1$, then we set
$$
W(\phi)=W(\psi_0)\cup W(\psi_1)\cup\{\ttt{true,false}\}.
$$
Finally, we set
$$
W(\neg\psi)=W(\psi)\cup\{\ttt{true,false}\}.
$$
Again, inspection of the relevant clauses in the semantics of guards
shows that the conclusion of the lemma holds.
\end{proof}

Finally, we prove the corresponding result for rules.

\begin{la}
For every rule $R$, there is a bounded exploration witness $W(R)$.
\end{la}

\begin{proof}
We define $W(R)$ by recursion on $R$ as follows.

If $R$ is an update rule $f(t_1,\dots,t_n):=t_0$, then
$$
W(R)=\bigcup_{i=0}^nW(t_i).
$$

If $R$ is $\ttt{issue}f(t_1,\dots,t_n)$, then
$$
W(R)=\bigcup_{i=1}^nW(t_i).
$$

If $R$ is \ttt{fail} then $W(R)$ is empty.

If $R$ is a conditional rule $\ttt{if\ }\phi\ttt{\ then\ }R_0\ttt{\
  else\ }R_1$, then
$$
W(R)=W(\phi)\cup W(R_0)\cup W(R_1)\cup\{\ttt{true,false}\}.
$$

If $R$ is a parallel combination $\ttt{do\ in\ parallel\
}R_1,\dots,R_k$ then
$$
W(R)=\bigcup_{i=1}^kW(R_i).
$$

That $W(R)$ serves as a bounded exploration witness for $R$ is proved
by induction on $R$.  Every case of the inductive proof is trivial in
view of the previous lemmas and the definition of the semantics of
rules.
\end{proof}

\section{Algorithms are Equivalent to ASMs}   
\label{sec:thesis}

In this section, we shall prove the Abstract State Machine Thesis for
interactive, small-step algorithms.  That is, we shall prove that
every algorithm (as defined in \cite[Section~3]{ga1}) is equivalent
(as defined in \cite[Section~4]{ga1}) to an ASM (as in
Definition~\ref{asm-def}).

Throughout this section, we assume that we are given an interactive,
small-step algorithm $A$.  By definition, it has a set \scr S of
states, a set \scr I of initial states, a finite vocabulary
$\U$, a finite set $\Lambda$ of labels, causality relations
$\vdash_X$, sets $\scr F_X$ of final histories, subsets $\scr F_X^+$
and $\scr F_X^-$ of successful and failing final histories, and
update sets $\DD(X,\xi)$.  Here and throughout this section, $X$
ranges over states and $\xi$ over histories for $X$.  Furthermore,
$A$ has, by the Bounded Work Postulate of \cite{ga1} and its
corollaries, a bound $B$ for the number and lengths of the queries
issued in any state under any attainable history, and it has a
bounded exploration witness $W$. Since $W$ retains the property of
being a bounded exploration witness if more $\U$-terms are
added to it, we may assume that $W$ is closed under subterms and
contains \ttt{true,\ false}, and some variable.

To define an ASM equivalent to $A$, we must specify, according to
Definition~\ref{asm-def},
\begin{itemize}
\item its vocabulary,
\item its set of labels,
\item its external vocabulary,
\item its program,
\item its template assignment,
\item its set of states, and its set of initial states.
\end{itemize}

\subsection{Vocabulary, labels, states}

Some of these specifications are obvious, because the definition of
equivalence requires that the vocabulary, the labels, the states, and
the initial states be the same for our ASM as they are for the given
algorithm $A$. It remains to define the external vocabulary, the
template assignment, and the program.

Before proceeding, we note that Definition~\ref{asm-def} requires \scr
S and \scr I to be closed under isomorphisms and requires \scr S to be
closed under the transitions of the ASM.  The first of these
requirements is satisfied by our choice of \scr S and \scr I because
$A$ satisfies the Isomorphism Postulate.  That the second requirement
is also satisfied will be clear once we verify that the update sets
and therefore the transition functions of $A$ and of our ASM agree (at
least on successful final histories), for the Step Postulate ensures
that \scr S is closed under the transitions of $A$.

\subsection{External vocabulary and templates}
\label{sub:extvoc}

To define the external vocabulary $\E$ and the template assignment for
our ASM, we consider all templates, of length at most $B$, for the
given set $\Lambda$ of labels, in which the placeholders $\#i$ occur
in order.  (Recall that $B$ is an upper bound on the lengths of
queries issued by algorithm $A$ in arbitrary states for arbitrary
attainable histories.)  These templates, which we call \emph{standard
templates}, can be equivalently described as the tuples obtained by
taking any initial segment of the list $\#1,\#2,\dots$ of placeholders
and inserting elements of $\Lambda$ into such a tuple, while keeping
the total length of the tuple $\leq B$.  We note that any potential
query of length $\leq B$ (over any state) is obtained from a unique
standard template by substituting elements of the state for the
placeholders.  We define the external vocabulary $\E$ and the template
assignment simultaneously by putting into $\E$ one function symbol $f$
for each standard template and writing $\hat f$ for the standard
template associated to $f$.  Define an external function symbol $f$ to
be $n$-ary if $\hat f$ is a template for $n$-ary functions.

\begin{rmk}
For many algorithms, the external vocabulary defined here is larger
than necessary; many symbols in $\E$ won't occur in the program $\Pi$.
One can, of course, discard such superfluous symbols once $\Pi$ is
defined.  We chose the present definition of $\E$ in order to make it
independent of the more complicated considerations involved in
defining $\Pi$.  
\qed\end{rmk}

\begin{rmk}
We have not specified --- nor is there any need to specify --- exactly
what entities should serve as the external function symbols $f$
associated to templates $\hat f$.  The simplest choice mathematically
would be to take the function symbols to be the standard templates
themselves, but even with this choice, which would make $\hat f=f$, it
would seem worthwhile to maintain the notational distinction between
$f$, to be thought of as a function symbol, and $\hat f$, to be
thought of as a template.
\qed\end{rmk}

\subsection{Critical elements, critical terms, agreement}

The preceding discussion completes the easy part of the definition
of our ASM; the hard part that remains is to define the program
$\Pi$. Looking at the characterization in Lemma 4.3 of \cite{ga1},
we find that we have (trivially) satisfied the first requirement for
the equivalence of our ASM and the given $A$ (agreement as to
states, initial states, vocabulary, and labels), and that we must
construct $\Pi$ so as to satisfy the remaining three requirements
(agreement as to queries issued, finality, success, failure, and
updates).  Notice that these three requirements refer only to
histories that are attainable for both algorithms.  This means that,
in constructing $\Pi$, we can safely ignore what $A$ does with
unattainable histories.

As in the proofs of the ASM thesis for other classes of algorithms in
\cite{seqth, parth, oa3}, we use the bounded exploration witness to gain
enough control over the behavior of the algorithm $A$ to match it with
an ASM.  The first step in this process is the following lemma, whose
basic idea goes back to \cite{seqth}.

\begin{df}
  Let $X$ be a state and $\xi$ a history for it.  An element $a\in X$
  is \emph{critical} for $X$ and $\xi$ if there is a term $t\in W$ and
  there are values in $\ran{\ans\xi}$ for the variables in $t$ such
  that the resulting value for $t$ is $a$.
\qed\end{df}

\begin{la}[Critical Elements]    \label{crit-la}
Let $X$ be a state, $\xi$ a coherent history for it, and $a$ an
element of $X$. Assume that one of the following holds.
\begin{ls}
  \item There is a query $q$ such that $\xi\vdash_Xq$ and $a$ is one
  of the components of the tuple $q$.
  \item There is an update $\sq{f,\sq{b_1,\dots,b_n},c}\in\DD(X,\xi)$
  such that $a$ is one of the $b_i$'s or $c$.
\end{ls}
Then $a$ is critical for $X$ and $\xi$.
\end{la}

\begin{proof}
The proof is very similar to the one in \cite[Propositions~5.23 and
5.24]{oa1}, so we shall be rather brief here.
We may assume, as an
induction hypothesis, that the lemma holds when $\xi$ is replaced with
any proper initial segment of $\xi$.  (This is legitimate because
initial segments inherit coherence from $\xi$.)
Because $\xi$ is coherent, every query in its domain is caused by
some proper initial segment.  So all components in $X$ of such a query
are critical for that initial segment and therefore also critical for
$\xi$.

Assume that $a$ is not critical for $X$ and $\xi$; we shall show that
neither of the two hypotheses about $a$ can hold.

Form a new state $X'$, isomorphic to $X$, by replacing $a$ by a new
element $a'$.  Since $a$ is not critical, it is neither a component of
a query in \dom{\ans\xi} nor an element of \ran{\ans\xi}.  Thus $\xi$ is a
history for $X'$ as well as for $X$.  Using again the assumption that
$a$ is not critical, one finds that $(X,\xi)$ and $(X',\xi)$ agree on
$W$.  As $W$ is a bounded exploration witness for $A$, and as $a$ is
obviously neither a component of a query caused by $\xi$ over $X'$ nor
a component in an update in $\DD(X',\xi)$ (because $a\notin X'$), it
follows that $a$ is neither a component of a query caused by $\xi$
over $X$ nor a component in an update in $\DD(X,\xi)$.
\end{proof}

The construction of our ASM will be similar to that in
\cite[Section~5]{oa3}, but some additional work will be needed to
take into account the timing information in histories and the
possibility of incomplete but final histories.

The role played by element tags (or e-tags) and query tags (or q-tags)
in \cite{oa3} will now be played by ASM-terms, i.e., variable-free
terms over the vocabulary $\U \cup \E$. Some of these terms, those
with outermost function-symbol in $\E$, will, by
Definition~\ref{term-val-def}, obtain two kinds of possible values:
the ordinary value which is an element of the state, and also a
query-value, which is a potential query.

\begin{df}[Critical Terms] Recall that an ASM-term is a
  closed term of the vocabulary $\U \cup \E$.
  \begin{ls}
    \item A critical  term of \emph{level} 0 is a closed term in the
    bounded exploration witness $W$.
    \item If $t_1,\ldots,t_k$ are critical terms with maximal level
    $n$ and $f$ is a $k$-ary function symbol in $\E$, then
    $f(t_1,\dots,t_k)$ is a critical q-term of level $n+1$.
    \item If $t \in W$ contains exactly the variables $x_1,\ldots,x_k$
    and if $t_1,\ldots,t_k$ are critical q-terms with maximal level $n$,
    then the result of substituting $t_i$ for $x_i$ in $t$, $i =
    1,\ldots,k$, is a critical term of level $n$.
    \item By a \emph{critical  term} we mean a critical  term of some level.
  \end{ls}
\qed\end{df}

Because we arranged for $W$ to contain a variable, the third clause of
the definition implies that every critical q-term is a critical term
(of the same level), so our terminology is consistent.

Since $W$ and the external vocabulary $\E$ are finite, there are
only finitely many critical terms of any one level.

Notice that, although they are obtained from the $\U$-terms
in $W$, our critical terms are ASM-terms.  That is, they
contain no variables, but they can contain external function
symbols.

The values of ASM-terms, including in particular critical terms, for
a given state $X$ and history $\xi$, as well as the query-values of
q-terms, were defined in Definition~\ref{term-val-def}.

Recall also that, according to Lemma~\ref{mon-t}, any term that has
a value in state $X$ with respect to an initial segment of $\xi$
will have the same value with respect to $\xi$ itself, and that the
same holds of query values. The next lemma records some related
facts for future reference. Recall that two pairs $(X,\xi)$ of a
state and history are said to agree on $W$ if the two histories are
the same and every term in $W$ gets the same values (if any) in both
states when the variables are given values in the range of the
history.

\begin{la}[Invariance of Values]  \label{tag-val-invar}
\mbox{}
  \begin{ls}
    \item If $i:X\cong Y$ is an isomorphism, $\xi$ is a history for
    $X$, and $z$ is any ASM-term, then $i(\val zX\xi)=\val zY{i(\xi)}$.
    If $z$ is a q-term, then also $i(\qval zX\xi)=\qval zY{i(\xi)}$.
    \item If $(X,\xi)$ and $(Y,\xi)$ agree on $W$, then $\val
    zX\xi=\val zY\xi$ for all critical terms $z$, and
    $\qval zX\xi=\qval zY\xi$ for all critical q-terms $z$.
  \end{ls}
\end{la}

\begin{proof}
  The first assertion is proved by induction on terms, using the
  Isomorphism Postulate.  The second is proved by induction on
  critical terms and critical q-terms, using the facts that all
  critical terms are, by definition, in the bounded exploration
  witness $W$ and that the same history $\xi$ is used on both sides of
  the claimed equations.
\end{proof}

\begin{rmk}
An approximation to the intuition behind critical terms is that critical
terms of level $n$ represent (for a state $X$ and history $\xi$) the
elements of $X$ and the queries that can play a role in the computation of
our algorithm $A$ during the first $n$ rounds or phases of its interaction
with the environment.  This is based on the intuition that the bounded
exploration witness $W$ represents all the things the algorithm can do,
with the environment's replies, to focus its attention on elements of $X$.
At first, before receiving any information from the environment (indeed,
before even issuing any queries), the algorithm can focus only on the
values of closed terms from $W$, i.e., the values of critical terms of
level 0.  Using these, it can formulate and issue queries; these will be
query-values of q-terms of level 1.  Once some replies are received to
those queries, the algorithm can focus on the values of non-closed terms
from $W$ with the replies as values for the variables.  The replies are
the values for the q-terms of level 1 that denote the issued queries, and
so the elements to which the algorithm now pays attention are the values
of critical terms of level $\leq1$.  Using them, it assembles and issues
queries, query-values of q-terms of level $\leq2$. The replies, used as
values of the variables in terms from $W$, give the new elements to which
the algorithm can pay attention, and these are the values of critical
terms of level $\leq2$.  The process continues similarly for later rounds
of the interaction with the environment and correspondingly higher level
terms.

One should, however, be careful not to assume too much about the
connection between levels of critical terms and rounds of
interaction. It is possible for a critical term $t$ of level 1 to
acquire a value only after many rounds of interaction, if, for
example, the history happens to answer many other queries, one after
the other, before finally getting to one that is needed for evaluating
$t$.  It is also possible for a critical term of high level to acquire
a value earlier than its level would suggest. Consider, for example, a
critical term of the form $f(f(f(0)))$, where $f$ is an external
function symbol and 0 is a constant symbol from $\U$.  If the
history $\xi$ contains just one reply, giving the query $\hat f[0_X]$
the value $0_X$, then this suffices to give $f(f(f(0)))$ the value
$0_X$.

The following lemma formalizes the part of this intuitive explanation
that we shall need later.
\qed\end{rmk}

\begin{la}[Critical Terms Suffice]  \label{tags-suff}
Let $X$ be a state, $\xi$ an attainable history for it, and $n$ the
length of $\xi$.
\begin{ls}
  \item Every query in \dom{\ans\xi}\ is the query-value (for $X$ and $\xi$) of
some critical q-term of level $\leq n$.
   \item Every element of \ran{\ans\xi} is the value (for $X$ and
   $\xi$) of some critical q-term of level $\leq n$.
   \item Every critical element for $X$ and $\xi$ is the value (for
   $X$ and $\xi$) of a critical term of level $\leq n$.
   \item Every query in $\Issued_X(\xi)$ is the query-value (for $X$ and
$\xi$) of some critical q-term of level $\leq n+1$.
\end{ls}
\end{la}

\begin{proof}
We proceed by induction on the length $n$ of the history $\xi$. As
$\xi$ is coherent, any query in its domain is issued by a proper
initial segment $\eta\init\xi$.  So, by induction hypothesis
(applied to the last clause), such a query is the query-value of a
q-term of level $\leq\text{length}(\eta)+1 \le n$. This proves the
first assertion of the lemma.

The second follows, because, if a query in \dom{\ans\xi}\ is the
query-value of a q-term of level $\le n$, then the reply given by
$\xi$ is the value of the same term.

For the third assertion, consider any critical element, say the value
of a term $t\in W$ when the variables of $t$ are given certain values
in \ran{\ans\xi}.  By the second assertion already proved, these
values of the variables are also the values of certain critical
q-terms of level $\leq n$. Substituting these terms for the variables
in $t$, we obtain a critical term of level $\leq n$ whose value is the
given critical element.

For the final assertion, consider any query issued by $\xi$.  It has
length at most $B$ (by our choice of $B$), so it is obtained by
substituting elements of $X$ for the placeholders in some standard
template.  That is, it has the form $\hat f[a_1,\dots,a_k]$ for some
external function symbol $f$ and some elements $a_i\in X$.  By
Lemma~\ref{crit-la}, each $a_i$ is critical with respect to $X$ and
$\xi$.  By the third assertion already proved, each $a_i$ is the value
of some critical term $t_i$ of level $\leq n$.  Then our query $\hat
f[a_1,\dots,a_k]$ is the query-value of the critical q-term
$f(t_1,\dots,t_k)$ of level $\leq n+1$.
\end{proof}

As indicated earlier, we can confine our attention to attainable
histories.  The lengths of these are bounded by $B$, and so we
may, by the lemma just proved, confine our attention to critical
terms of level at most $B$.  In particular, only a finite set of
critical terms will be under consideration.

We have the following partial converse to the last statement of
Invariance of Values Lemma~\ref{tag-val-invar}.  We abbreviate the
phrase ``pair consisting of a state and an attainable history for it''
as ``attainable pair.''

\begin{la}[Agreement]\label{lemma:agree}
Let $(X,\xi), (Y,\xi)$ be attainable pairs with $\xi$ of length $n$.
If they agree as to the values of all critical terms of level
$\le n$, then they agree on $W$.
\end{la}

\begin{proof}
Note that in the assumption we didn't mention agreement as to
query-values. But $(X,\xi)$ and $(Y,\xi)$ will agree as to
query-values of critical q-terms of level $n$ as soon as they agree as
to the values of critical terms of levels $<n$.

Let $t \in W$. We need to prove that it takes the same value in
$(X,\xi)$ and $(Y,\xi)$ when all variables in $t$ are given values
in $\ran{\ans\xi}$. But values in $\ran{\ans\xi}$ are, by the
Critical Terms Suffice Lemma~\ref{tags-suff}, the values of some
critical q-terms of level $\le n$. Substituting these terms for the
variables in $t$ gives us, by definition, a critical term of level
$\le n$, where by assumption $(X,\xi)$ and $(Y,\xi)$ agree.
\end{proof}

\subsection{Descriptions, similarity}

The following definitions are intended to capture all the information
about a state and history that can be relevant to the execution of our
algorithm $A$.  That they succeed will be the content of the subsequent
discussion and lemmas.

\begin{df}
Let $(X,\xi)$ be an attainable pair.  Let $n$ be
the length of $\xi$.  (Recall that $n$ is finite and in fact $\leq
B$.)  Define the \emph{truncation} $\xi-$ of $\xi$ to be the initial
segment of $\xi$ of length $n-1$ (undefined if $n=0$).  The
\emph{description} $\delta(X,\xi)$ of $X$ and $\xi$ is the Kleene
conjunction of the following guards:
\begin{ls}
\item all equations $s=t$ and negated equations $\neg(s=t)$ that have
  value \ttt{true} in $(X,\xi)$, where $s$ and $t$ are critical terms
  of level $\leq n$, and
\item all timing inequalities $(u\prec v)$ and $(u\preceq v)$ that
  have value \ttt{true} in $(X,\xi)$, where $u$ and $v$ are critical
  q-terms of level $\le n$, and where \qval vX\xi\ exists and is in
  $\Issued_X(\xi-)$.
\end{ls}
\qed\end{df}

Some comments may help to clarify the last clause here, about
timing inequalities.  First, recall that the strict inequality
$(u\prec v)$ is merely an abbreviation of $\neg(v\preceq u)$.

Second, although we explicitly require only $v$ to have a query-value
in $\Issued_X(\xi-)$,
the same requirement for $u$ is included in the
requirement that $(u\preceq v)$ or $(u\prec v)$ is true.  Indeed,
inspection of the definition of the semantics of timing guards (in
Definition~\ref{guard-sem-def}) shows that the q-term $u$ must have a
value, and this is possible only if $u$ has a query-value in
\dom{\ans\xi}.  Since $\xi$ is coherent, it follows that \qval uX\xi\
must be in $\Issued_X(\xi-)$.

Third, if $n=0$ then $\xi-$ is undefined, and as a result
$\delta(X,\xi)$ contains no timing inequalities.

Our definition of the description of $X$ and $\xi$ is not complete on
the syntactic level, for it does not specify the order or
parenthesization of the conjuncts in the Kleene conjunction.  That is,
it is complete only up to associativity and commutativity of $\kand$.
The reader is invited to supply any desired syntactic precision; it
will never be used. The choice of order and parenthesization of
conjuncts makes no semantic difference; the Kleene conjunction and
disjunction are commutative and associative as far as truth values and
issued queries are concerned.

We shall sometimes refer to descriptions of attainable pairs as
\emph{attainable descriptions}, even though ``attainable'' is
redundant here because the descriptions have been defined only for
attainable pairs.

The following lemma and its corollary provide useful information about
the q-terms occurring in a description.

\begin{la}
Let $(X,\xi)$ be an attainable pair, $n\geq1$ the length of $\xi$, and
$v$ a q-term.  The following are equivalent.
\begin{lsnum}
\item $v$ occurs in $\delta(X,\xi)$.
\item $v$ occurs as one side of a timing inequality in
  $\delta(X,\xi)$.
\item $v$ is a critical q-term of level $\leq n$ and it has a
  query-value $\qval vX{\xi-}$ that is in $\Issued_X(\xi-)$.
\end{lsnum}
\end{la}

\begin{proof}
Since the implication from (2) to (1) is trivial, we prove that (3)
implies (2) and that (1) implies (3).

Suppose first that (3) holds.  Let $q$ be any query in the last
equivalence class of the preorder in $\xi$.  As $\xi$ is attainable,
$q\in\Issued_X(\xi-)$.  Also, by Lemma~\ref{tags-suff}, $q=\qval
uX\xi$ for some critical q-term $u$
of level $\leq n$.  Because $q$ is in the
last equivalence class with respect to $\xi$, \val uX\xi\ exists but
\val uX{\xi-} does not.  Now if $\qval vX{\xi-}$, which is also $\qval
vX\xi$, is in \dom{\ans\xi}, then \val vX\xi\ exists and so
$\delta(X,\xi)$ contains the conjunct $(v\preceq u)$.  Otherwise, \val
vX\xi\ does not exist, and so $\delta(X,\xi)$ contains the conjunct
$(u\prec v)$.  In either case, (2) holds.

Finally, we assume (1) and deduce (3).  Inspection of the definition
of descriptions reveals that any q-term $v$ that occurs in
$\delta(X,\xi)$ must be a sub-q-term either of some critical term of
level $\leq n$ that has a value with respect to $(X,\xi)$ or of some
critical q-term of level $\leq n$ that either has a value with respect
to $(X,\xi)$ or at least has a query-value that is issued with respect
to $(X,\xi-)$.  In any case it follows, thanks to the attainability
(and in particular the coherence) of $\xi$, that (3) holds.
\end{proof}

\begin{coro}
  The q-terms that occur in the description of an attainable pair
  $(X,\xi)$ depend only on $X$ and $\xi-$, not on the last equivalence
  class in the preorder of \dom{\ans\xi}.
\end{coro}

\begin{proof}
  Immediate from the third of the equivalent statements in the lemma.
\end{proof}

Clearly, $\delta(X,\xi)$ is a guard, and
$\val{\delta(X,\xi)}X\xi=\ttt{true}$.  The next lemma shows that
descriptions are invariant under two important equivalence relations
on attainable pairs $(X,\xi)$.

\begin{la}[Invariance of Descriptions] \label{invar-descr}
Let $(X,\xi)$ be an attainable pair.
\begin{itemize}
\item If $(Y,\xi)$ is an attainable pair (with the same $\xi$)
agreeing with $(X,\xi)$ on $W$, then they have the same
descriptions.
\item If $(Y,\eta)$ is an attainable pair isomorphic to $(X,\xi)$,
then they have the same descriptions.
\end{itemize}
\end{la}

\begin{proof} To see that the first statement is true, use the second
  clause of Invariance of Values Lemma~\ref{tag-val-invar} to
  establish that the same critical terms occur in
$\delta(X,\xi)$ and $\delta(Y,\xi)$ in the same roles. To see that the
second statement is true, use the first clause of the same lemma.
\end{proof}

Thus, each of agreement and isomorphism is a sufficient condition for
similarity in the sense of the following definition.  We shall see
later, in Corollary~\ref{agree-cor}, that the composition of agreement
and isomorphism is not only sufficient but also necessary for
similarity.

\begin{df} Two attainable pairs are \emph{similar} if they have the
  same descriptions.
\qed\end{df}

 The next lemma describes the other states and histories in which
$\delta(X,\xi)$ is true, and thus leads to a characterization of
similar attainable pairs.

\begin{la}   \label{agree-la}
Let $(X,\xi)$ and $(Y,\eta)$ be attainable pairs.  Suppose
$\delta(X,\xi)$ has value \ttt{true} in $(Y,\eta)$.  Then
\begin{itemize}
\item the length of $\eta$ is at least the length of $\xi$;
\item there is an attainable pair $(Z,\eta')$ isomorphic to $(X,\xi)$,
  such that $\eta'$ is an initial segment of $\eta$ and $(Z,\eta')$
  agrees with $(Y,\eta')$ on $W$.
\end{itemize}
\end{la}

In other words, any $(Y,\eta)$ that satisfies the description of
$(X,\xi)$ can be obtained from $(X,\xi)$ by the following three-step
process.  First, replace $(X,\xi)$ by an isomorphic copy $(Z,\eta')$.
Second, leaving the history $\eta'$ unchanged, replace $Z$ by a new
state $Y$ but maintain agreement on the bounded exploration witness
$W$.  Third, extend the history $\eta'$ by adding new items strictly
after the ones in $\eta'$, so that $\eta'$ is an initial segment of
the resulting $\eta$.

Notice that, by virtue of the isomorphism of $(X,\xi)$ and
$(Z,\eta')$, we can describe $\eta'$ more specifically as the initial
segment of $\eta$ of the same length as $\xi$.

\begin{proof}
We proceed by induction on the length $n$ of the history $\xi$.

\smallskip\noindent\textbf{Length: $\eta$ is not shorter than $\xi$.}
Choose one query from each of the $n$ equivalence classes in
\dom{\ans\xi}, say $q_j$ from the $j\th$ equivalence class.  Letting
$\xi\restr j$ denote the initial segment of $\xi$ of length $j$, and
applying Lemma~\ref{tags-suff}, we express each $q_j$ as the query-value,
with respect to $(X,\xi\restr j)$, of some critical q-term $u_j$ of level
$\leq j$.  Thus, $u_j$ has a value $\dot\xi(q_j)$ with respect to
$\xi\restr j$ but not with respect to $\xi\restr(j-1)$.  Thus,
$\delta(X,\xi)$ includes the conjuncts $(u_j\prec u_{j+1})$ for
$j=1,2,\dots,n-1$ and also the conjunct $u_n=u_n$.  So these conjuncts
must also be true in $(Y,\eta)$, which means that $\eta$ has length at
least $n$.

\medskip\noindent\textbf{Construction of $(Z,\eta')$.} Our next
step will be to define a certain isomorphic copy $(Z,\eta')$ of
$(X,\xi)$.  Afterward, we shall verify that $\eta'$ has the other
properties required.

 We may assume, by replacing $(X,\xi)$ with an isomorphic copy if
necessary, that $X$ is disjoint from $Y$.  Next, obtain an isomorphic
copy $Z$ of $X$ as follows.  For each critical term $t$ of level $\leq
n$, if \val tX\xi\ exists, then remove this element from $X$ and put
in its place the element \val tY\eta\ of $Y$.  To see that this makes
sense, we must observe two things.  First, the equation $t=t$ is one
of the conjuncts in $\delta(X,\xi)$ and is therefore true for $Y$ and
$\eta$.  Thus, the replacement element \val tY\eta\ exists.  Second,
if the same element of $X$ is also \val{t'}X\xi\ for another critical
term $t'$ of level $\leq n$, then the equation $t'=t$ is a conjunct in
$\delta(X,\xi)$ and is therefore true for $Y$ and $\eta$.  Thus,
\val{t'}Y\eta=\val tY\eta, which means that the replacement element is
uniquely defined.

Let $i$ be the obvious isomorphism from $X$ to $Z$, sending each of
the replaced elements \val tX\xi\ to its replacement \val tY\eta\ and
sending all the other elements of $X$ to themselves.  Let
$\eta'=i(\xi)$; this is the history for $Z$ obtained by applying $i$ to
all components from $X$ in the queries in \dom{\ans\xi}\ and to all the
replies in \ran{\ans\xi}.  Because of the isomorphism, it is clear
that $(Z,\eta')$ is, like $(X,\xi)$, an attainable pair and that
$\eta'$ has the same length $n$ as $\xi$.

\medskip\noindent\textbf{Values: $\dot\eta'$ is a subfunction of $\dot
\eta$.}  Consider any query $q=\hat f[a_1,\dots,a_k]\in\dom{\ans\xi}$
and its reply $b=\ans\xi(q)$.  Thus, $i(q)\in \dom{i(\ans\xi)}$, and
$i(\ans\xi)(i(q))=i(b)$.  Furthermore, every element of
\dom{i(\ans\xi)} is $i(q)$ for some such $q$.  By
Lemma~\ref{tags-suff}, all the $a_j$ are values in $(X,\xi)$ of
certain critical terms $t_j$ of level $<n$, and so $b$ is the value of
the critical term $f(t_1,\dots,t_k)$ of level $\leq n$.  In forming
$Z$, we replaced the elements $a_j$ by the values $i(a_j)$ of the
$t_j$'s in $(Y,\eta)$, and we replaced $b$ by $i(b)$, the value in
$(Y,\eta)$ of $f(t_1,\dots,t_k)$.  But this last value is, by
definition, the result of applying $\ans\eta$ to the query that is the
query-value of $f(t_1,\dots,t_k)$, namely the query $\hat
f[i(a_1),\dots,i(a_k)]=i(q)$.  That is, $i(b)=\ans\eta(i(q))$.  This
shows that, whenever $i(\ans\xi)$ maps a query $i(q)$ to a reply
$i(b)$, then so does $\ans\eta$; in other words, $\ans\eta'$ is a
subfunction of $\ans\eta$.

\medskip\noindent\textbf{Order: $\le_{\eta'}$ is a sub-preorder of
$\le_\eta$.} We next show that the preordering of $\eta'$ agrees with
that of $\eta$.  Consider an arbitrary $q\in\dom{\ans\xi}$, and
suppose it is in the $j\th$ equivalence class with respect to the
preorder given by $\xi$.  So, as $\xi$ is coherent,
$q\in\Issued_X(\xi\restr(j-1))$, and so, by the last part of
Lemma~\ref{tags-suff}, we have a critical q-term $u$ of level $\leq j$
such that $q=\qval uX{\xi\restr(j-1)}$.  Note that \val
uX{\xi\restr(j-1)} does not exist, because
$q\notin\dom{\ans\xi\restr(j-1)}$.

We wish to apply the induction hypothesis to $(X,\xi\restr(j-1))$.  To
do so, we observe that $\delta(X,\xi\restr(j-1))$ is a subconjunction
of $\delta(X,\xi)$ and is therefore true in $(Y,\eta)$.  So we can
apply the induction hypothesis and find that $(X,\xi\restr(j-1))$ is
isomorphic to an attainable pair that agrees with
$(Y,\eta\restr(j-1))$.  By Lemma~\ref{tag-val-invar}, $u$ has a
query-value but no value in $(Y,\eta\restr(j-1))$.  Inspection of the
definitions shows that its query-value is $i(q)$.

If $j<n$, i.e., if $q\in\dom{\xi-}$, then we can also apply the
induction hypothesis to $(X,\xi\restr j)$, in which $u$ has a value.
We conclude that $u$ has a value in $(Y,\eta\restr j)$.  Since it had
a query-value but no value in $(Y,\eta\restr(j-1))$, we conclude that
its query-value, $i(q)$ must be in exactly the $j\th$ equivalence
class with respect to $\eta$.

If, on the other hand, $j=n$, i.e., if $q$ is in the last equivalence
class with respect to $\xi$, then this last application of the
induction hypothesis is not available.  Nevertheless, since
$q\in\dom{\ans\xi}$, we know that $u$ has a value in $(X,\xi)$, so
$\delta(X,\xi)$ contains the conjunct $u=u$, so this conjunct is true
also in $(Y,\eta)$, and so $u$ has a value in $(Y,\eta)$.  This means
that $i(q)$, the query-value of $u$, is in $\dom{\ans\eta}$.  We saw
earlier that it is not in $\dom{\ans\eta\restr(j-1)}$, where now
$j=n$.  So $i(q)$ is in the $n\th$ equivalence class or later with
respect to $\eta$.

What we have proved so far suffices to establish that if $q<_\xi q'$
then $i(q)<_\eta i(q')$ and that the same holds for non-strict
inequalities except in the case that both $q$ and $q'$ are in the last
equivalence class with respect to $\xi$.  In this exceptional case, we
know that $i(q)$ is the query-value, already existing in
$(Y,\eta\restr(n-1))$, of $u$ (as above), yet $u$ has no value in
$(Y,\eta\restr(n-1))$. This means that the smallest $m$ for which \val
uY{\eta\restr m} exists is the $m$ such that $i(q)$ is in the $m\th$
equivalence class with respect to $\eta$.  Repeating the argument with
an analogously defined q-term $u'$ for $q'$, and using the fact that
$\delta(X,\xi)$ contains the conjuncts $(u\preceq u')$ and $(u'\preceq
u)$, which means that these conjuncts are also true in $(Y,\eta)$, we
find that $i(q)$ and $i(q')$ are in the same equivalence class with
respect to $\eta$.

This completes the proof that $\eta'$ --- including both the answer
function and the pre-order --- is the restriction of $\eta$ to some
subset of its domain.  In fact, we have shown more, namely that, for
$j<n$, $i$ maps the $j\th$ equivalence class with respect to $\xi$
into the $j\th$ equivalence class with respect to $\eta$, and that it
maps the last ($n\th$) equivalence class with respect to $\xi$ into a
single equivalence class --- possibly the $n\th$ and possibly later
--- with repect to $\eta$.

The next step is to show that $\eta'=i(\xi)$ is an initial segment of
$\eta$.  This will imply that, in the preceding summary of what was
already proved, both occurrences of ``into'' can be improved to
``onto'' and ``possibly the $n\th$ and possibly later'' can be
improved to ``the $n\th$''.

\medskip\noindent\textbf{Initial segment: $\eta'$ is an initial
segment of $\eta$.}  Suppose, toward a contradiction, that
\dom{\ans\eta'} is not an initial segment of \dom{\ans\eta} (with
respect to $\leq_\eta$).  So there exist some
$q\in\dom{\ans\eta}-\dom{\ans\eta'}$ and some $q'\in\dom{\ans\xi}$
(and thus $i(q')\in\dom{\ans\eta'}$) such that $q\le_\eta i(q')$.
Among all such pairs $q,q'$, fix one for which $q$ occurs as early as
possible in the preorder $\le_\eta$.  Since $q'\in\dom{\ans\xi}$, we
can fix a critical q-term $u'$ of level $\leq n$ with
$\qval{u'}X\xi=q'$ and thus, by definition of $i$,
$\qval{u'}Y\eta=i(q')$.  We record for future reference that, since
$\qval{u'}X\xi\in\dom{\ans\xi}$, $u'$ has a value with respect to
$\xi$.

Consider the initial segment of $\eta$ up to but not including $q$. By
what we have already proved (and our choice of $q$ as the earliest
possible), it is $i(\zeta)$ for some proper initial segment $\zeta$ of
$\xi$ --- proper because it doesn't contain $q'$.  In particular,
$\zeta$ has length at most $n-1$, and so we know, by induction
hypothesis, that the lemma is true with $\zeta$ in place of $\xi$.
(As before, the lemma can be applied because $\delta(X,\zeta)$ is a
subconjunction of $\delta(X,\xi)$, which is true in $(Y,\eta)$.)  So
we conclude that $(X,\zeta)$ is isomorphic to an attainable pair that
agrees on $W$ with $(Y,i(\zeta))$.

Since $i(\zeta)$ is the initial segment of $\eta$ ending just before
$q$, and since $\eta$ is a coherent history, we know that
$q\in\Issued_Y(i(\zeta))$.  By the Critical Terms Suffice
Lemma~\ref{tags-suff}, $q$ is the query-value in $(Y,i(\zeta))$, and
therefore also in $(Y,\eta)$, of some critical q-term $u$ of level
$\le n$. Thanks to the isomorphism between $(X,\zeta)$ and an
attainable pair agreeing with $(Y,i(\zeta))$, we have that $u$ also
has a query-value, say $q''$, in $(X,\zeta)$ and this value is in
$\Issued_X(\zeta)$ and, a fortiori, in $\Issued_X(\xi)$.  By
definition of $i$, $i(q'')=q$. As $q$ was chosen outside
$\dom{i(\ans\xi)}=i(\dom{\ans\xi})$, it follows that
$q''\notin\dom{\ans\xi}$. From this and $q'\in\dom{\ans\xi}$, we
conclude that $(u'\prec u)$ is one of the conjuncts in $\delta(X,\xi)$
and is therefore true in $(Y,\eta)$. Since $u$ has a value in the
initial segment of $\eta$ up to and including $q$ (one equivalence
class beyond $i(\zeta)$), we infer that $u'$ must have a value in
$(Y,i(\zeta))$.  That means that the query-value of $u'$, namely
$i(q')$ must be in the domain of $i(\zeta)$, i.e., $i(q')<_\eta q$.
This contradicts the original choice of $q$ and $q'$, and this
contradiction completes the proof that $\eta'$ is an initial segment
of $\eta$.

\medskip\noindent\textbf{Agreement: $(Z,\eta')$ and $(Y,\eta')$ agree
  on $W$.}
It remains to prove that the attainable pairs $(Z,\eta')$ and
$(Y,\eta')$ agree on $W$.  We prove this in three steps.

First, we show that, if $z$ is any critical term of level $\leq
n$, then
$$
i(\val zX\xi)=\val zY{\eta'}.
$$
This is almost the definition of $i$, which says that $i(\val
zX\xi)=\val zY\eta$.  Our task is to replace $\eta$ on the right
side of this equation with $\eta'$.  That is, we must show that,
if \val zX\xi\ exists (and therefore \val zY\eta\ exists), then
\val zY{\eta'} exists, because then we shall have $\val
zY{\eta'}=\val zY\eta$ by the monotonicity of values.  We proceed
by induction on the level of $z$.
The only non-trivial case, i.e., the only case where changing
$\eta$ to $\eta'$ could matter, is the case that $z$ is a q-term.
The possibility that we must exclude is that $\qval zY\eta$ (which
is also \qval zY{\eta'} as the induction hypothesis applies to the
arguments of $z$) is in the domain of $\eta$ but not in the domain
of $\eta'$. But $\qval zX\xi$ exists and is in the domain of $\xi$
(because \val zX\xi\ exists), and its image under $i$ is, by
definition of $i$, \qval zY\eta.  So this image is in the domain
of $i(\xi)=\eta'$, as desired.

Second, we observe that, since $i$ is an isomorphism from $X$ to $Z$
and sends $\xi$ to $\eta'$, we have $i(\val zX\xi)=\val zZ{\eta'}$.
Combining this with the result established in the preceding paragraph,
we have
$$
\val zZ{\eta'}=\val zY{\eta'}
$$
for all critical terms $z$ of level $\leq n$.

Finally, an application of the Agreement Lemma~\ref{lemma:agree}
completes the proof that $(Z,\eta')$ and $(Y,\eta')$ agree on $W$.
\end{proof}

\begin{coro}[Factorization]    \label{agree-cor}
Let $(X,\xi)$ and $(Y,\eta)$ be similar attainable pairs.   Then
there is a state $Z$ such that $\eta$ is an attainable history for
$Z$, $(Z,\eta)$ agrees with $(Y,\eta)$ on $W$, and $(Z,\eta)$ is
isomorphic to $(X,\xi)$.
\end{coro}

\begin{proof}
We can apply Lemma~\ref{agree-la} to $(X,\xi)$ and $(Y,\eta)$ in
either order, since each satisfies the other's description.  Thus
$\xi$ and $\eta$ have the same length, and the $\eta'$ of the lemma is
simply $\eta$.  The rest of the corollary is contained in the lemma.
\end{proof}

\begin{coro}[Similarity Suffices]    \label{descr-suff}
Let $(X,\xi)$ and $(Y,\eta)$ be similar attainable pairs.   Let
$n$ be the length of $\xi$ (and of $\eta$, by
Corollary~\ref{agree-cor}). Then
\begin{ls}
\item If $u$ is a q-term of level $\leq n+1$ and $\xi\vdash_X\qval
uX\xi$
  then $\eta\vdash_Y\qval uY\eta$.
\item If $\xi$ is in $\scr F_X^+$ or $\scr F_X^-$, then $\eta$ is in
  $\scr F_Y^+$ or $\scr F_Y^-$, respectively.
\item If $\DD(X,\xi)$ contains an update \sq{f,\sq{a_1,\dots,a_k},a_0}
  where each $a_i$ is $\val{t_i}X\xi$ for a critical term $t_i$ of level $\leq
  n$, then $\DD(Y,\eta)$ contains the update
  \sq{f,\sq{a'_1,\dots,a'_k},a'_0} where each $a'_i$ is
  $\val{t_i}Y\eta$.
\end{ls}
\end{coro}

\begin{proof}
Apply the preceding corollary to get $Z$ such that $(Z,\eta)$
agrees with $(Y,\eta)$ on $W$ and is isomorphic to $(X,\xi)$.
Because of the agreement on the bounded exploration witness $W$,
we have all the desired conclusions with $(Z,\eta)$ in place of
$(X,\xi)$.  To complete the proof, we can replace $(Z,\eta)$ with
$(X,\xi)$, thanks to the Isomorphism Postulate and the fact that
isomorphisms respect evaluation of terms.
\end{proof}

We shall also need the notion of a \emph{successor} of an attainable
description.  This corresponds to adjoining one new equivalence class
at the end of a history, while leaving the state unchanged. That is,
$\delta(X,\xi)$ is a successor of $\delta(X,\xi-)$, and
$\delta(X,\xi-)$ is the \emph{predecessor} of $\delta(X,\xi)$.

\begin{rmk}
To avoid possible confusion, we emphasize that a successor of
$\delta(X,\eta)$ need not be of the form $\delta(X,\xi)$ with
$\xi-=\eta$.  It could instead be of the form $\delta(Y,\xi)$ for some
other pair $(Y,\xi)$ such that $(Y,\xi-)$ is similar to $(X,\eta)$,
and there might be no way to extend $\eta$ so as to obtain similarity
with $(Y,\xi)$.  For a simple example, suppose the bounded exploration
witness $W$ contains only \ttt{true}, \ttt{false}, \ttt{undef}, and a
variable.  Let $X$ be a structure containing only the three elements
that are the values of \ttt{true}, \ttt{false}, and \ttt{undef}, and
let $Y$ be like $X$ but with one additional element $a$.  Suppose
further that the algorithm is such that a single query $q$, say
\sq{\ttt{true}}, is caused by the empty history $\emp$ in every state.
Then $(X,\emp)$ and $(Y,\emp)$ agree on $W$, and $Y$ admits an
attainable history $\xi$ with $\dom{\ans\xi}=\{q\}$ and with
$\ans\xi(q)=a$.  Then, since $\xi-=\emp$, we have that $\delta(Y,\xi)$
is a successor of $\delta(Y,\emp)=\delta(X,\emp)$.  But there is no
history $\zeta$ for $X$ such that $\delta(Y,\xi)=\delta(X,\zeta)$; such
a $\zeta$ would have to map $q$ to a value distinct from \ttt{true},
\ttt{false}, and \ttt{undef}, and $X$ has no such element.
\qed\end{rmk}

The use of the definite article in ``the predecessor'' is justified by
the following observation, showing that $\delta(X,\xi-)$ is completely
determined by $\delta(X,\xi)$.  Thus, ``predecessor'' is a
well-defined operation on attainable descriptions of non-zero length.
Of course the situation is quite different for successors; one
description can have many successors because there are in general many
ways to extend an attainable history by appending one more equivalence
class.

\begin{coro}
  Let $(X,\xi)$ and $(Y,\eta)$ be similar attainable pairs,
  and assume the (common) length of $\xi$ and $\eta$ is
  not zero.  Then $\delta(X,\xi-)=\delta(Y,\eta-)$.
\end{coro}

\begin{proof}
By Corollary~\ref{agree-cor}, we have an isomorphism
$i:(X,\xi)\cong(Z,\eta)$ such that $(Z,\eta)$ agrees with $(Y,\eta)$
on $W$.  Since the isomorphism $i$ must, in particular, respect the
pre-orderings, it follows immediately that $i$ is also an isomorphism
from $(X,\xi-)$ to $(Z,\eta-)$.  From the definition of agreement, it
follows immediately that $(Z,\eta-)$ and $(Y,\eta-)$ agree on $W$.
Thus, by Lemma~\ref{invar-descr}, $(X,\xi-)$ and $(Y,\eta-)$ are
similar.
\end{proof}

The following information about successors will be useful when we
verify that the ASM that we produce is equivalent to the given
algorithm $A$.

\begin{la}   \label{successor}
  Suppose $(X,\xi)$ is an attainable pair and $\delta'$ is an
  attainable description that is a successor of $\delta(X,\xi)$.
  Then $\delta'=\delta(Y,\eta)$ for some attainable pair $(Y,\eta)$
  such that
  \begin{ls}
    \item $\xi=\eta-$,
    \item $(X,\xi)$ and $(Y,\xi)$ agree on $W$.
  \end{ls}
\end{la}

\begin{proof}
By definition of successor, we have an attainable pair $(Z,\theta)$
such that $\delta'=\delta(Z,\theta)$ and
$\delta(X,\xi)=\delta(Z,\theta-)$.  This last equality implies, by
Corollary~\ref{agree-cor}, that $(X,\xi)$ and $(Y,\xi)$ agree on $W$
for some attainable pair $(Y,\xi)$ isomorphic to $(Z,\theta-)$.  Use
the isomorphism to transport $\theta$ to an attainable history $\eta$
for $Y$.  Then $\delta'=\delta(Z,\theta)=\delta(Y,\eta)$ because of
the isomorphism, and $\eta-$ is the image, under the isomorphism, of
$\theta-$, i.e., $\eta-=\xi$.
\end{proof}

\subsection{The ASM program}

We are now ready to describe the ASM program that will simulate our
given algorithm $A$.  Its structure will be a nested alternation of
conditionals and parallel combinations, with updates, issue rules, and
\ttt{fail} as the innermost constituents.  The guards of the
conditional subrules will be attainable descriptions.  Recall that the
critical terms involved in attainable descriptions all have levels
$\leq B$, and there are only finitely many such terms and therefore
only finitely many attainable descriptions.  An attainable description
$\delta(X,\xi)$ will be said to have \emph{depth} equal to the length
of $\xi$.  Lemma~\ref{agree-la} ensures that this depth depends only
on the description $\delta(X,\xi)$, not on the particular attainable
pair $(X,\xi)$ from which it is obtained.  Notice that the definition
of descriptions immediately implies that any critical term occurring
in a description has level $\leq$ the depth of the description.

We construct the program $\Pi$ for an ASM equivalent to the given
algorithm $A$ as follows.  $\Pi$ is a parallel combination, with one
component for each attainable description $\delta$ of depth zero. We
describe the component associated to $\delta$ under the assumption
that $\delta$ is not final, by which we mean that, in the attainable
pairs $(X,\xi)$ with description $\delta$, the history $\xi$ is not
final; we shall return later to the final case.  (Recall
that, by Corollary~\ref{descr-suff}, whether $\xi$ is final in $X$
depends only on the description $\delta(X,\xi)$, so our case
distinction here is unambiguous.)

The component associated to a non-final $\delta$ is a conditional rule
of the form $\ttt{if\ }\delta\ttt{\ then\ }R_\delta$, i.e., a
conditional whose guard is $\delta$ itself.  The body $R_\delta$ is a
parallel combination, with one component for each successor $\delta'$
of $\delta$.

When $\delta'$ is not final, the associated component is
a conditional rule $\ttt{if\ }\delta'\ttt{\ then\ }R_{\delta'}$.  The
body $R_{\delta'}$ here is a parallel combination, with one component
for each successor $\delta''$ of $\delta'$.

Continue in this manner until a final description $\eps$ is reached.
Since the depth increases by one when we pass from a description to a
successor, and since all attainable histories have length (i.e., the
depth of their descriptions) at most $B$, we will have reached final
descriptions after at most $B$ iterations of the procedure.  The
component associated to a final description $\eps=\delta(X,\xi)$ is
$\ttt{if\ }\eps\ttt{\ then\ }R_\eps\ttt{\ endif}$, where $R_\eps$ is
the parallel combination of the following:
\begin{ls}
\item \ttt{fail} if $\xi\in\scr F_X^-$,
\item $\ttt{issue}\ u$ if $u$ is a q-term of level at most one more
than the length of $\xi$ (that is, the depth of $\eps$) and
$\xi\vdash_X\qval uX\xi$, and
\item $f(t_1,\dots,t_k):=t_0$ if the $t_i$ are critical terms of
level at most
  the length of $\xi$ and they have values $a_i=\val{t_i}X\xi$ such
  that $\sq{f,\sq{a_1,\dots,a_k},a_0}\in\DD(X,\xi)$.
\end{ls}
It is important to note that, although the attainable pair
$(X,\xi)$ was used in the specification of these components, they
actually depend only on the description $\eps$, by
Corollary~\ref{descr-suff}. This completes the definition of the
program $\Pi$.

\begin{rmk}
  As in previous work on the ASM thesis, this program $\Pi$ is
  designed specifically for the proof of the thesis.  That is, it
  works in complete generality and it admits a fairly simple, uniform
  construction.  For practical programming of specific algorithms,
  there will normally be ASM programs far simpler than the one
  produced by our general method.
\end{rmk}

\subsection{Equivalence}

It remains to show that the ASM defined by $\Pi$ is equivalent to the
given algorithm $A$.  For brevity, we sometimes refer to this ASM as
simply $\Pi$.

\begin{thm}   \label{main-thm}
The ASM defined by $\Pi$ together with  \scr S, \scr I,
  $\U$, $\Lambda$, $\E$, and the template assignment of
  subsection~\ref{sub:extvoc} is equivalent to algorithm $A$.
\end{thm}

\begin{proof}
 Referring to Lemma 4.3 of \cite{ga1}, we see that it suffices to
show the following, for every pair $(X,\xi)$ that is attainable for
both the algorithm $A$ and our ASM.
\begin{lsnum}
\item $\Issued_X(\xi)$ is the same for our ASM as for $A$.
\item If $\xi$ is in $\scr F_X^+$ or $\scr F_X^-$ with respect to one
  of $A$ and our ASM,
  then the same is true with respect to the other.
\item If $\xi\in\scr F_X^+$, then $\DD(X,\xi)$ is the same with
  respect to our ASM and with respect to $A$.
\end{lsnum}

Consider, therefore, an attainable pair $(X,\xi)$ (with respect to
$A$) and the behavior of our ASM in this pair.

Let $n$ be the length of $\xi$, and for each $m\leq n$ let $\xi\restr
m$ be the initial segment of $\xi$ of length $m$.  According to
Lemma~\ref{agree-la}, the only attainable descriptions satisfied by
$(X,\xi)$ are those of the form $\delta(X,\xi\restr m)$, one of each
depth $m\leq n$.

\medskip\noindent\textbf{Issuing Queries.}

We begin by analyzing the queries issued by our ASM in state $X$
with history $\xi$.  (Parts of this analysis will be useful again
later, when we analyze finality, success, failure, and updates.)
For readability, our analysis will be phrased in terms of the ASM
performing various actions, such as issuing queries or passing
control to a branch of a conditional rule.  Of course, this could
be rewritten more formally in terms of the detailed  semantics of
ASMs, but the formalization seems to entail more costs, both for
the reader and for the authors, than benefits.

The ASM acting in state $X$ with history $\xi$ begins, since $\Pi$
is a parallel combination, by executing all the components
associated with attainable descriptions of depth 0.  Recall that
these components are conditional rules whose guards are the
descriptions themselves. These descriptions contain only critical
terms of depth 0, so there are no external function symbols here.
Therefore, no queries result from the evaluation of the guards. By
Lemma~\ref{agree-la} the ASM finds exactly one of the guards to be
true, namely $\delta(X,\xi\restr0)$, and it proceeds to execute
the body $R_{\delta(X,\xi\restr0)}$ of this conditional rule.

Let us suppose, temporarily, that $n>0$, so, as $\xi$ is attainable,
$\xi\restr0$ is not final.  (We shall return to the other case later.)
So $R_{\delta(X,\xi\restr0)}$ is a parallel combination, and our ASM
proceeds to execute its components.  These are conditionals, whose
guards $\delta'$ are the successors of $\delta(X,\xi\restr0)$.  So
these guards are $\delta(Y,\eta)$ for attainable pairs $(Y,\eta)$ as in
Lemma~\ref{successor}.  In particular, $\eta$ has length 1 and
$\eta-=\xi\restr0$.  (This last equation is redundant as both sides
are histories of length 0, but we include it to match what will occur
in later parts of our analysis.)  Inspection of the definition of
descriptions shows that every query issued during the evaluation of
such a guard is also issued by the algorithm $A$ operating in the
attainable pair $(Y,\eta-)=(Y,\xi\restr0)$.  Since $(Y,\xi\restr0)$
agrees with $(X,\xi\restr0)$ on $W$, these are queries issued by $A$
in $(X,\xi\restr0)$.

The converse also holds.  If a query $q$ is issued by $A$ in
$(X,\xi\restr0)$, then there is an attainable history $\eta$ for
$X$ in which $q$ is in the first and only equivalence class of
$\dom{\ans\eta}$; simply define $\eta$ to give $q$ an arbitrary reply
and to do nothing more.  By Lemma~\ref{tags-suff}, $q$ is the
query-value of some q-term $u$ of level 1, and therefore
$\delta(X,\eta)$ contains the conjunct $u=u$.  Thus, in evaluating
the guard $\delta(X,\eta)$, our ASM will issue $q$.

Having evaluated the guards of depth 1, our ASM finds, according to
Lemma~\ref{agree-la}, that exactly one of them is true, namely
$\delta(X,\xi\restr1)$, so it proceeds to evaluate the corresponding
body $R_{\delta(X,\xi\restr1)}$.  Let us suppose, temporarily, that
$n>1$, so, as $\xi$ is attainable, $\xi\restr1$ is not final.  So
$R_{\delta(X,\xi\restr1)}$ is a parallel combination, and our ASM
proceeds to execute its components.  These are conditionals, whose
guards $\delta'$ are the successors of $\delta(X,\xi\restr1)$.  So
these guards are $\delta(Y,\eta)$ for attainable pairs $(Y,\eta)$ as
in Lemma~\ref{successor}.  In particular, $\eta$ has length 2 and
$\eta-=\xi\restr1$.  Inspection of the definition of descriptions
shows that every query issued during the evaluation of such a guard is
also issued by the algorithm $A$ operating in the attainable pair
$(Y,\eta-)=(Y,\xi\restr1)$.  Since $(Y,\xi\restr1)$ agrees with
$(X,\xi\restr1)$ on $W$, these are queries issued by $A$ in
$(X,\xi\restr1)$.

The converse also holds.  If a query $q$ is issued by $A$ in
$(X,\xi\restr1)$, but not already in $(X,\xi\restr0)$, then there is
an attainable history $\eta$ for $X$, which has $\xi\restr1$ as an
initial segment, and in which $q$ is in the second and
last equivalence class of $\dom{\ans\eta}$; simply define $\eta$ by
extending $\xi\restr1$ to give $q$ an arbitrary reply, in a new,
second equivalence class, and to do nothing more.
By Lemma~\ref{tags-suff}, $q$ is the query-value of some critical term
$u$ of level 2, and therefore $\delta(X,\eta)$ contains the conjunct
$u=u$.  Thus, in evaluating the guard $\delta(X,\eta)$, our ASM will
issue $q$.

The reader should, at this point, experience d\'ej\`a vu, since the
argument we have just given concerning the behavior of our ASM while
executing $R_{\delta(X,\xi\restr1)}$ is exactly parallel to the
previous argument concerning $R_{\delta(X,\xi\restr0)}$.  The same
pattern continues as long as the depths of the guards are $<n$ so that
we have not arrived at a final history.

Consider now what happens when the ASM evaluates
$R_{\delta(X,\xi\restr n)}=R_{\delta(X,\xi)}$.  If the history $\xi$
is not final, then the same argument as before shows that the ASM will
issue, while evaluating the guards of the components of
$R_{\delta(X,\xi)}$, the same queries as the original algorithm $A$.
Furthermore, the ASM will find none of the guards here to be true, for
these guards are descriptions of depth $n+1$ and can, by
Lemma~\ref{agree-la}, be satisfied only with histories of length at
least $n+1$.  So the execution of the ASM produces no additional
queries beyond those that we have already shown to agree with those
produced by $A$.

There remains the situation that $\xi$ is final for $A$ and $X$.  In
this case, the components of $R_{\delta(X,\xi)}$ are no longer
conditional rules, the evaluation of whose guards causes the
appropriate queries to be issued by the ASM.  Rather, the components
are issue rules, updates, or \ttt{fail}.  Only the issue rules here
will result in new queries; the queries involved in evaluating the
terms in update rules and in the issue rules have already been issued
during the evaluation of guards.  And the issue rules are chosen
precisely to issue the queries that $A$ would issue in $(X,\xi)$.

This completes the proof that our ASM and $A$ agree as to issuing
queries.  They therefore agree as to which histories are coherent.

\medskip\noindent\textbf{Finality, Success, and Failure.}
We next consider which histories are declared final by our ASM.
Suppose first that $\xi$ is final for $A$ in $X$.  Then, as the
preceding analysis of the ASM's behavior shows, the ASM will,
after evaluating a lot of guards, find itself executing
$R_{\delta(X,\xi)}$, which is a parallel combination of issue
rules, update rules, or \ttt{fail}.  The subterms of any update
rules here will already have been evaluated during the evaluation
of the guards, so $\xi$ is final for these update rules.  The same
goes for the issue rules; their subterms have already been
evaluated, and so $\xi$ is final.  Any history is final for
\ttt{fail}.  Thus $\xi$ is final for all the components of
$R_{\delta(X,\xi)}$ and is therefore final for $R_{\delta(X,\xi)}$
itself.  From the definition of the semantics of parallel combinations
and conditional rules, it follows that $\xi$ is also final for $\Pi$,
as required.

Now suppose that $\xi$ is (attainable but) not final for $A$ in
state $X$.  There will be some queries that have been issued by
$A$ but not answered, i.e., that are in
$\Issued_X(\xi)-\dom{\ans\xi}=\text{Pending}_X(\xi)$, for otherwise
$\xi$ would be complete and attainable and therefore, by the Step
Postulate, final.  So our ASM will issue some queries whose
answers are needed for the evaluation of the guards of some
components of $R_{\delta(X,\xi)}$, but whose answers are not in
$\xi$.  Therefore, $\xi$ is not a final history for the ASM in
state $X$.  This completes the proof that our ASM agrees with $A$
as to finality of histories.

We check next that a final history $\xi$ succeeds or fails for our
ASM according to whether it succeeds or fails for $A$. It fails
for our ASM if and only if, after evaluating all the guards and
while executing $R_{\delta(X,\xi)}$, it encounters either
\ttt{fail} or clashing updates (see the definition of failure for
parallel combinations).  By definition of our ASM, it encounters
\ttt{fail} if and only if $A$ fails in $(X,\xi)$. Furthermore, it
will not encounter clashing updates unless $A$ fails, because, as
we shall see below, it encounters exactly the updates produced by
$A$, and these cannot, by the Step Postulate, clash unless $A$
fails.

\medskip\noindent\textbf{Updates.} To complete the proof, we have to
 check what updates our ASM encounters.  Our
construction of $\Pi$ is such that update rules are encountered
only in the subrules $R_{\delta(X,\xi)}$ for final histories
$\xi$. Furthermore, these update subrules are chosen to match the
updates performed by $A$.  So our ASM and $A$ produce the same
updates in any final history.

This completes the verification that our ASM is equivalent to the
given algorithm $A$.
\end{proof}

\section{Concluding Remarks}

Theorem~\ref{main-thm} establishes the ASM thesis for small-step,
interactive algorithms, as defined by the postulates of \cite{ga1}.
This completes the program of proving the ASM thesis in the small-step
case.  The case of parallel algorithms, without intrastep interaction,
was treated in \cite{parth}, but the task of combining intrastep
interaction with parallelism remains for future work.  Beyond that,
there is the task of treating distributed algorithms.  

The ASM syntax and semantics presented in Sections~\ref{sec:asm} and
\ref{asm-sem} serve to describe just what has to be added to the
traditional ASM syntax and semantics of \cite{G103} in order to
accommodate non-ordinary interaction.  Essentially, one needs timing
guards and the Kleene connectives.  It remains to be seen whether
these additions will also suffice to model all interactive parallel
algorithms.

\vskip-40 pt

\begin{thebibliography}{99}

\bibitem{parth} Andreas Blass and Yuri Gurevich, ``Abstract state machines
capture parallel algorithms,'' \emph{ACM Trans. Computational Logic},
4 (2003) 578--651.

\bibitem{oa1}
Andreas Blass and Yuri Gurevich, ``Ordinary Interactive Small-Step
Algorithms, I,'' \emph{ACM Trans. Computational Logic}, 7 (2006)
363--419. 

\bibitem{oa2}
Andreas Blass and Yuri Gurevich, ``Ordinary Interactive Small-Step
Algorithms, II,'' \emph{ACM Trans. Computational Logic}, to
appear.

\bibitem{oa3}
Andreas Blass and Yuri Gurevich, ``Ordinary Interactive Small-Step
Algorithms, III,'' \emph{ACM Trans. Computational Logic}, to
appear.

\bibitem{ga1} Andreas Blass, Yuri Gurevich, Dean Rosenzweig, and
Benjamin Rossman, ``Interactive Small-Step Computation I:
Axiomatization,'' in preparation.

\bibitem{composite} Andreas Blass, Yuri Gurevich, Dean Rosenzweig, and
Benjamin Rossman, ``Composite interactive algorithms'' (tentative
title), in preparation.

\bibitem{G64.5}
Yuri Gurevich, ``A new thesis,'' Abstract 85T-68-203,
\emph{Amer. Math. Soc. Abstracts} 6 (August, 1985) p.317.

\bibitem{G92}
Yuri Gurevich,
``Evolving Algebras: An Introductory Tutorial,''
Bull. EATCS 43 (February 1991), 264--284.
Reprinted with slight revisions in
\emph{Current Trends in Theoretical Computer Science: Essays and Tutorials},
G. Rozenberg and A. Salomaa (editors),
World Scientific, 1993, 266-292

\bibitem{G103}
Yuri Gurevich,
``Evolving algebra 1993: Lipari guide,''
in \emph{Specification and Validation Methods},
E. B\"orger (editor), Oxford Univ. Press (1995) 9--36.

\bibitem{seqth}
Yuri Gurevich, ``Sequential abstract state machines capture sequential
algorithms,'' \emph{ACM Trans. Computational Logic} 1 (2000) 77--111.

\bibitem{web} James K. Huggins, ASM Michigan web page,
\href{http://www.eecs.umich.edu/gasm}
    {\ttt{http://www.eecs.umich.edu/gasm}}

\bibitem{Kleene} Stephen Cole Kleene, \emph{Introduction to
  Metamathematics},  Van Nostrand, 1952.

\end{thebibliography}
\end{document}